\documentclass[fleqn,usenatbib]{mnras}

\usepackage{newtxtext,newtxmath}

\usepackage[T1]{fontenc}

\DeclareRobustCommand{\VAN}[3]{#2}
\let\VANthebibliography\thebibliography
\def\thebibliography{\DeclareRobustCommand{\VAN}[3]{##3}\VANthebibliography}
\usepackage{amstext}
\usepackage{amsmath}
\usepackage{cases}
\usepackage{url}
\usepackage[]{times,refname,amsmath,bm}
\bibpunct{(}{)}{;}{a}{}{,}
\usepackage{tabularx}
\usepackage{graphicx,epsfig,color,latexsym}
\usepackage[subpreambles=true]{standalone}
\usepackage{import}

\newcommand{\bea}{\begin{eqnarray}}
\newcommand{\eea}{\end{eqnarray}}
\newcommand{\be}{\begin{equation}}
\newcommand{\ee}{\end{equation}}
\newcommand{\rund}[1]{\left(#1\right)}
\newcommand{\vc}[1]{\mbox{\boldmath $#1$}}

\newcommand{\eck}[1]{\left[ #1 \right]}

\usepackage{ulem}

\usepackage{academicons}
\usepackage{xcolor}
\newcommand{\orcid}[1]{\href{https://orcid.org/#1}{\textcolor[HTML]{A6CE39}{\aiOrcid}}}

\title[Binary microlensing with plasma]{Binary microlensing with plasma environment -- Star and planet}

\author[Sun, Er \& Tsupko]{Jiarui Sun$^{1}$\thanks{Email:1013412198@qq.com}, Xinzhong Er$^2$\thanks{Email: phioen@163.com} and Oleg Yu. Tsupko$^{3}\thanks{Email: tsupko@iki.rssi.ru; tsupkooleg@gmail.com}$\\
$^1$ Department of Astronomy, Yunnan University, Kunming, 650500 P.R.China\\
$^2$ South-Western Institute for Astronomy Research, Yunnan University, Kunming, 650500 P.R.China\\
$^{3}$ Space Research Institute of Russian Academy of Sciences, Profsoyuznaya 84/32, Moscow 117997, Russia
}

\date{Accepted XXX. Received YYY; in original form ZZZ}

\pubyear{2022}

\begin{document}
\label{firstpage}
\pagerange{\pageref{firstpage}--\pageref{lastpage}}
\maketitle

\begin{abstract}
Galactic microlensing has been widely used to study the star and planet. The stellar wind plays an important role in the formation, environment and habitability of the planet. In this work we study a binary microlensing system including the stellar wind, i.e. a star with plasma environment plus a planet. Plasma surrounding the main lens causes chromatic deflection of the light rays, in addition to the gravitational one. As a result, such a lensing system can generate complicated caustics which depends on the different lensing parameters. In this work we study the magnification curves for different traces of the background source and compare the transitions of the formation of ``hill and hole'' in the magnification curves. We find that the plasma will cause extra caustic, shrink the central caustics generated by the star and push the caustic by the planet outwards. Observations and modelling of binary microlensing curves with taking plasma effect into account can provide a potential method to study plasma environment of the stars. In case of a high plasma density of the stellar wind, the plasma lensing effects will be observable in the sub-mm band.
\end{abstract}


\section{Introduction}

In gravitational microlensing, the mass of the lens is small. This implies that the lensing effects, such as image separation of multiple images and the image distortions are difficult to detect. The microlensing in the Local Group (or Galactic microlensing) refers to the temporal brightening of a background star due to intervening compact objects \citep[e.g.,][]{2012RAA....12..947M}. Since the first discovery of microlensing event by \citet{1993Natur.365..621A, 1993AcA....43..289U}, this field has made enormous progress. In particular, the gravitational microlensing has been developed into a powerful technique in the detection of exoplanets \citep{1991ApJ...374L..37M}. More and more planets have been found through large scale optical surveys, e.g., OGLE \citep[][]{2015AcA....65....1U}; MACHO  \citep[][]{2000ApJ...541..734A} and KMT  \citep[][]{2016JKAS...49...37K}. By means of various other techniques, several thousand exoplanets have been confirmed to date\footnote{https://exoplanetarchive.ipac.caltech.edu}. Microlensing not only provides a complementary to these methods in searching for exoplanets, but also detects planets around all types of stars
\citep[e.g.][]{2006glsw.book.....S, 2012RAA....12..947M, 2012ARA&A..50..411G}, and even planets around binary system (which can be called 'triple system') \citep[e.g.][]{2022MNRAS.516.1704K}.

The microlensing light curves are usually achromatic, because the gravitational deflection of light is independent of wavelength. However, there can be reasons leading to chromatic effects in microlensing. First, there exists the well-known phenomenon of chromatic microlensing in strong lensed quasars \citep[e.g.][]{1986A&A...166...36K, 1991AJ....102..864W}. The source can have different size at different wavelength, thus will be lensed with different magnifications. Another possibility of chromatic effects is attributed to the propagation of light in the plasma \citep[e.g., review by][]{2017Univ....3...57B}. Chromatic refractive deflection can be caused by the ionised gas in the interstellar medium (ISM), or by more concentrated plasma near astrophysical objects.

The light propagation in presence of medium 
in the gravitational field
was considered by \citet{1960rgt..book.....S}, and further investigated for deflection in presence of plasma by \citet{2000rofp.book.....P}; for related discussion, see \citet{1975A&A....44..389B, 2021PhRvD.103j4019T}. Gravitational lensing in Schwarzschild metric with presence of non-homogeneous plasma has been extensively studied \citep[e.g.][]{1989grle.book.....B, 2009GrCo...15...20B, 2010MNRAS.404.1790B, BTreview15, 2013PhRvD..87l4009T,  2018PhRvD..97l4016C,2019PhRvD..99l4001C}. \citet{2014MNRAS.437.2180E,2022MNRAS.516.2218E} presented the effects due to plasma in the gravitational lensing applications, in particular the magnification and time delay effects in the strong lensing system.
In case of the lens being a compact object, such as a neutron star or a black hole, the surrounded plasma can significantly affect the rays, causing complex behaviour \citep[e.g.][]{2015MNRAS.451...17R, rogers17a, rogers17b, 2019MNRAS.484.2411K, 2022arXiv220614598B}. In the strong deflection case, the rays can make one or more revolutions around the lens, i.e. high-order images are formed \citep{2000PhRvD..62h4003V, 2001GReGr..33.1535B, 2004LRR.....7....9P}. They were investigated for the case with surrounding plasma in \citet{2013PhRvD..87l4009T}. Such extra deflection by plasma can also change the size and the shape of black hole shadow, which has been discussed by, e.g. \citet{2015PhRvD..92j4031P, 2017PhRvD..95j4003P, 2019PhRvD..99h4050Y, 2022JMP....63i2501B, 2022arXiv221105620B} and review \citet{2022PhR...947....1P}.

The stellar wind is highly ionised plasma and can reach a high density than the average value in the interstellar medium, e.g. the electron density near Earth is around $\sim 50$ electrons per centimetre cubic \citep{1958ApJ...128..664P}. The plasma near the star not only causes various magnification curves at different frequencies \citep{2020MNRAS.491.5636T}, it also provides a potential way to study the stellar wind of different types of stars. 
The emission from lots of sources is weak in radio bands, which causes the plasma lensing phenomena difficult to detect. However, since the electron density near the stars can be much higher than that in ISM, the plasma microlensing effects can be significant in high radio frequency or even in sub-millimetre band. 

The transformation of the microlensing curve from a single point lens to that with the presence of plasma has been first studied by \citet{2020MNRAS.491.5636T}. It has been shown that the refractive deflection in plasma can lead to a significant change in the shape of the curve, up to the disappearance of images due to strong divergence of rays. The authors called it 'hill-hole effect', implying that with sufficient plasma density (or observing at sufficiently long wavelength), the maximum in the centre of the microlensing curve ('hill') can be replaced by the minimum ('hole').

In this paper, we investigate how the refraction in plasma environment changes the microlensing curve in a more general case of a binary lens. We focus mainly on the case of a star surrounded by plasma and hosting a planet. In Sect.\,\ref{sec:formula}, we give short summary about the plasma lensing and introduce the binary microlensing with plasma environment. We study the magnification curves for a lensing system in Sect.\,\ref{sec:sim-curve}. In Sect.\,\ref{sec:caustics}, we present the evolving of the size of the caustics. The summary of our result is given in the end. In the appendix, we present the details of how we calculate the magnification, discuss the number of solutions, some caustics and critical condition.

\section{Equations of gravitational lensing in presence of plasma}
\label{sec:formula}
\subsection{Plasma deflection in gravitational lensing}

We briefly outline the formulae for gravitational and plasma deflection angles in this subsection. The basic of gravitational lensing can be found, e.g., in \citet{2006glsw.book.....S}. More details on the plasma deflection can be found in \citet{2020MNRAS.491.5636T}.
Let us consider a point mass lens surrounded by cold plasma with the refractive index
\begin{equation}
n_{pl}^2 = 1 - \frac{\omega_p^2}{\omega_0^2} \, ,    
\end{equation}
where $\omega_0$ is the observational frequency, and the plasma frequency $\omega_p$ is
\begin{equation}
\omega_p^2 = \frac{4 \pi e^2}{m_e} n_e \, .
\end{equation}
Here $n_e$ is the three dimensional number density of the electrons, $e$ is the electron charge, $m_e$ is the electron mass. 

With existing of plasma around a gravitational lens, the deflection angle $\hat{\alpha}$ will be caused by two contributions, the gravity and the plasma \citep[e.g.][]{2010MNRAS.404.1790B}:
\be
\hat{\alpha}(b) = \hat{\alpha}_{\rm grav}(b) +  \hat{\alpha}_{\rm pl}(b) \, ,
\label{eq:hat-alpha-grav}
\ee
where $b$ is the impact parameter of the light ray. The gravitational deflection by a point mass in vacuum is
\be \label{eq:hat-alpha-pl}
{\hat{\alpha}}_{\rm grav}(b) = \dfrac{4 GM}{c^2 b},
\ee
where $M$ is the mass of the lens. Since we only consider the microlensing cases, the point mass model of lens is sufficient in our study. And in this work, we only study the spherical symmetric distribution of plasma, i.e. the electron number density is a function of radius $r$ only, where $r=\sqrt{b^2+z^2}$. The deflection angle thus can be written as an integral alone the line of sight, i.e. $z-$direction \citep[e.g.][]{2010MNRAS.404.1790B, BTreview15}
\be
\hat{\alpha}_{\rm pl}(b) = \dfrac{4 \pi e^2}{m_e \omega_0^2} \int \dfrac{\partial n_e}{\partial b} dz \, .
\label{eq:plasma-alpha}
\ee
In our study, we limit to the small angle approximation, i.e., both angles (\ref{eq:hat-alpha-grav}) and (\ref{eq:hat-alpha-pl}) are small. In particular, it implies that $b \gg GM/c^2$ and $\omega_0 \gg \omega_p$.

Here, we assume that the total mass of the plasma is small compared with the lens, thus do not take into account of the mass of the plasma particles. The refractive deflection $\hat{\alpha}_{\rm pl}$ is caused by the density gradient of the plasma and usually to the opposite direction of that by the gravity of the lens star. With the convention $\hat{\alpha}_{\rm grav}>0$, the plasma deflection is negative in such case. 
Different density profiles have been used to study the plasma deflection, such as the Gaussian model \citep[][]{CleggFL1998,Tuntsov2016}. 

The density of free electrons near the Sun can be described by a power-law model. Thus we adopt that for our plasma lensing model in this study:
\be
n_e(r) = n_0 \rund{\frac{R_0}{r}}^h,
\ee
where $h>0$, and $n_0$ is a constant representing the electron density at a characteristic radius $r=R_0$ \citep[][]{1989grle.book.....B,BTreview15}. Eq.\,(\ref{eq:plasma-alpha}) becomes \citep[e.g.][]{2009GrCo...15...20B, BTreview15}
\be
\hat{\alpha}_{\rm pl}(b) = - \frac{4 \pi e^2 n_0}{m_e \omega_0^2}  \left( \frac{R_0}{b} \right)^h  \dfrac{\sqrt{\pi} \, \Gamma(\frac{h}{2}+\frac{1}{2})}{\Gamma(\frac{h}{2})} \, ,
\ee
where the Gamma function is
\be
\Gamma(x)=\int_0^{\infty} t^{x-1} e^{-t} dt \, .
\ee
This can be rewritten via wavelength $\omega_0 = 2 \pi c / \lambda$, where we neglect the difference between the group velocity of ray in plasma $n_{pl}c$ and vacuum velocity $c$ (since $\omega_p \ll \omega_0$). Using $r_e \equiv e^2/(m_e c^2)$, we write \citep[e.g.][]{1989grle.book.....B, er&rogers18}:
\be
\hat{\alpha}_{\rm pl} = - \lambda^2 r_e n_0   \left( \frac{R_0}{b} \right)^h  \dfrac{\Gamma(\frac{h}{2}+\frac{1}{2})}{\sqrt{\pi} \, \Gamma(\frac{h}{2})} \, .
\ee

\subsection{Lens equations for binary microlensing with plasma environment}
\label{sec:microlensing-plasma}

\begin{figure}
\centerline{\includegraphics[width=0.40\textwidth]{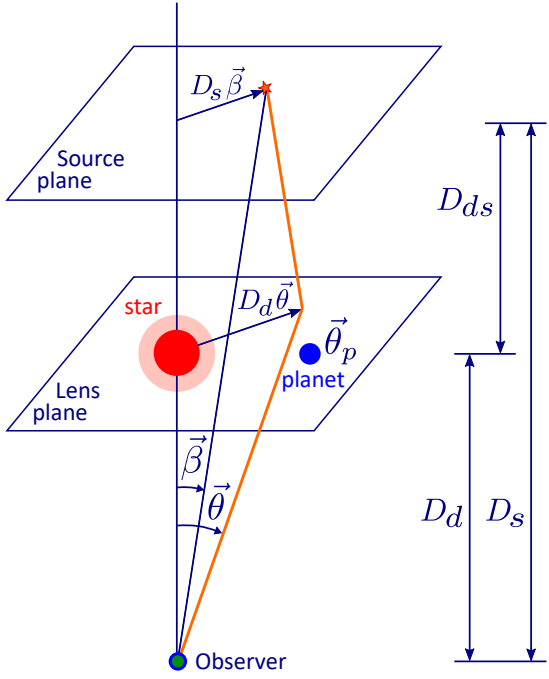}}
\caption{The geometry of the binary lensing. The big red dot represents the lens star with the pink shadow the plasma. The blue dot represents the planet.}
\label{fig:binary-plasma}
\end{figure}

It has been shown that plasma can affect the light curve of a microlensing event in case of point-mass lens \citep{2020MNRAS.491.5636T}. In this paper, we will investigate a more complicated case of binary system: a star surrounding with plasma, and hosting a planet as the lens. Since the problem is now not axially symmetric, we write all angles as two-dimensional vectors. 
We introduce the angular coordinates $\vc\theta=(\theta_1,\theta_2)$ in the lens plane, and those in the source plane $\vc\beta=(\beta_1,\beta_2)$. We use angular diameter distances between the lens and us, the source and us, and between the lens and the source as $D_d,D_s,D_{ds}$ respectively (Fig.\,\ref{fig:binary-plasma}). 
Then the scaled deflection angle $\vc{\alpha(\theta)}$ can be calculated by
\be
\vc{\alpha} = \frac{D_{ds}}{D_s} \vc{\hat{\alpha}}\, .
\ee
The lens equation can be written as %
\be
\vc{\beta} =\vc{\theta}-\vc{\alpha(\theta)}.
\ee
The image position $\theta$ is related to the impact parameter $\vc{b}=D_d \vc{\theta}$.
The magnification $\mu$ by a lens is inversely related to the Jacobian ${A}$ of the lens equation, such that:
\be
\mu=\frac{1}{{\rm det} A } \, , \;\; A= \begin{pmatrix}
\frac{\partial \beta_1}{\partial \theta_1}
&\frac{\partial \beta_1}{\partial \theta_2}\\ \\
\frac{\partial \beta_2}{\partial \theta_1}
&\frac{\partial \beta_2}{\partial \theta_2}
\label{eq:mu}
\end{pmatrix}
\ee
The curves along where the magnification diverges, i.e. ${\rm det} A=0$, are called critical curves on the lens plane. The corresponding curves on the source plane are the caustics.

We put the star at the origin of the coordinates for simplification, $\vc{\theta}=0$ in the lens plane. The position of the planet is at $\vc{\theta}_p$. The star is surrounded by plasma. With star and planet, the deflection angle by gravity can be given by 
\be
\vc{\alpha}_{\rm grav}=\frac{\theta^2_{E}\vc{\theta}}{|\vc{\theta}|^2}+\frac{\theta^2_{Ep}(\vc{\theta}-\vc{\theta_p})}{|\vc{\theta}-\vc{\theta}_p|^2},
\ee
where $\theta_{E},\, \theta_{Ep}$ is the Einstein radius of the star and of the planet respectively.

Based on the plasma model near the Sun \citep{1958ApJ...128..664P,1968ApJ...153..371C}, we adopt the power-law model of the electron density to the power of 2 as our plasma lens \citep{er&rogers18}. We define the angular scale length for this lens as 
\be
\theta_0 \equiv \rund{\lambda^2\dfrac{D_{ds}}{D_sD_d^2} \dfrac{r_e n_0 R_0^2}{\sqrt{\pi}} \dfrac{\Gamma(3/2)}{\Gamma(1)}}^{1/3}.
\label{eq:t0-ne}
\ee
Then the deflection angle due to plasma can be simplified to
\be
\vc{\alpha}_{\rm pl} = - \frac{\theta_0^3}{|\vc{\theta}|^2} \frac{\vc{\theta}}{|\vc{\theta}|} \, .
\ee
Then, the complete lens equation is
\be
\vc{\beta} = \vc{\theta}
- \frac{\theta_E^2 \vc{\theta}}{|\vc{\theta}|^2}
- \frac{\theta_{Ep}^2 (\vc{\theta} - \vc{\theta}_p) }{|\vc{\theta}-\vc{\theta}_p|^2} + \frac{\theta_0^3}{|\vc{\theta}|^2} \frac{\vc{\theta}}{|\vc{\theta}|} \, .
\label{eq:full-lenseq}
\ee

In order to simplify the mathematics, we substitute the lens equation by
\begin{equation}
\vc{y} = \frac{\vc{\beta}}{\theta_E} \, , \quad \vc{x} = \frac{\vc{\theta}}{\theta_E} \, , \quad \vc{x} = (x_1, x_2) \, , \quad \vc{y} = (y_1, y_2) \, ,
\end{equation}
Then the lens equation (\ref{eq:full-lenseq}) becomes
\be
\vc{y} = \vc{x} - \frac{\vc{x}}{|\vc{x}|^2}
- q \frac{\vc{x} - \vc{x}_p}{|\vc{x} - \vc{x}_p|^2}
+ \frac{B_0}{|\vc{x}|^2} \frac{\vc{x}}{|\vc{x}|} \, ,
\label{eq:reduced-lenseq}
\ee
where $x_p$ is the normalised separation between the star and planet $x_p=\theta_p/\theta_E$, $q$ is the mass ratio between the planet and the star
\be \label{eq:q-def}
q = \frac{M_p}{M} = \frac{\theta_{Ep}^2}{\theta_E^2}\, .
\ee
$B_0$ gives the strength of the plasma in the gravitational lens 
\be \label{eq:B0-def}
B_0=\frac{\theta_0^3}{\theta_E^3}.
\ee
In case of vacuum, $B_0=0$, the equation retrogresses to the lens equation of binary. We will present our figures in reduced coordinates $x,y$. Such a choice of variables is consistent with previous work in \cite{er&rogers18} where constant $\theta_0$ was used and in \cite{2020MNRAS.491.5636T} where parameter $B_0$ was introduced.

\section{Magnification curves of binary microlensing with plasma environment}
\label{sec:sim-curve}

In this Section, we numerically calculate microlensing magnification curves for the binary system under consideration: the star surrounded by plasma, and the planet. This system is described by the lens equation (\ref{eq:reduced-lenseq}). Normalised coordinates are used for calculation and for plotting figures: ($x_1, x_2$) for the lens plane and ($y_1, y_2$) for the source plane.

Numerical values of the most of lens parameters we adopt from \citet{1991ApJ...374L..37M}, see Tab.\,\ref{tab:lens_para}. For the simplification of numerical calculations, we take Einstein radius of the main lens as $\theta_E=1$ milli-arcsec. With our choice of the lens and the source distance (Tab.\,\ref{tab:lens_para}), the corresponding lens mass is about 0.984 Solar mass. The planet is chosen as $1/1000$ of the star mass, which is about the Jupiter mass. It means that the mass ratio is $q=0.001$. Lens separation in our normalised coordinates is $\vc{x_p}=(1.2,0)$.

\begin{table*}
\begin{tabular}{c|c|c|c|c|c|c|c|c|c}
$D_d$ &$D_s$ &$n_0$  &$R_0$ &$\theta_p$& $\theta_E$ &$\theta_{EP}$ & $q$ & $x_p$ & $B_0$ \\
\hline
4 kpc  &8 kpc & 5.6 cm$^{-3}$ &10 AU  &5 AU & 1 & $0.001^{1/2}$ & $0.001$ & 1.2 & $(0.1)^3; (0.7)^3; (0.8)^3$ \\
\hline
\end{tabular}
\caption{
The lens parameters, $\theta_E,\, \theta_{EP},$ are given in unit of milli-arcsec. $q,\, x_p,\, B_0$ are normalised by $\theta_E$.
The plasma model is taken from that near the Sun \citep{1958ApJ...128..664P}. }
\label{tab:lens_para}
\end{table*}

The effect of the plasma is completely determined by the parameter $B_0$,
which depends on the plasma density profile and the frequency of observation. Increasing the parameter $B_0$ corresponds to considering a denser plasma or decreasing the frequency of observation (increasing the wavelength).

\begin{figure}
\centerline{\includegraphics[width=6cm]{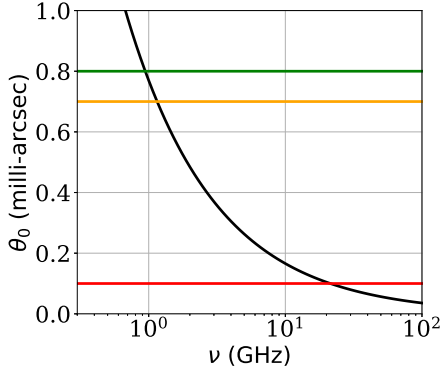}}
\caption{The relation between $\theta_0$ and observational frequency in our plasma lensing model (Eq.\,\ref{eq:t0-ne}).}
\label{fig:pl-t0}
\end{figure}

In our calculations, we consider different values of $B_0$. Since the plasma density profile is specified (Tab.\,\ref{tab:lens_para}), the different values of the plasma strength $B_0$ correspond to different observational frequency.
Generally, if we consider frequency from $0.65$ GHz to 100 GHz, this covers from low radio frequency to the milli-metre band. Such a range can give us the plasma lensing parameter $\theta_0$ from 1 to 0.03 milli-arcsec (Fig.\,\ref{fig:pl-t0}).
We take $\theta_0=0.1, 0.7, 0.8$ milli-arcsec. With $\theta_E=1$ milli-arcsec, this leads to $B_0=0.1^3,\, 0.7^3,$ and $0.8^3$. With values $n_0$, $R_0$, $D_d$, $D_s$ from Tab.\,\ref{tab:lens_para}, these three values of $B_0$ correspond to the observational frequencies roughly $21$ GHz, $1.1$ GHz and $0.93$ GHz respectively.
We show the angular scale length ($\theta_0$) as a function of the observational frequency in our plasma lensing model in Fig.\,\ref{fig:pl-t0}. The red, orange and green horizontal line show the three $\theta_0$ that we mentioned above. One can see that, as long as the frequency down to a few GHz, $\theta_0$ will be comparable with the Einstein radius of the star, and the plasma effects can be significant.

\begin{figure*}
\centerline{
\includegraphics[width=5.7cm]{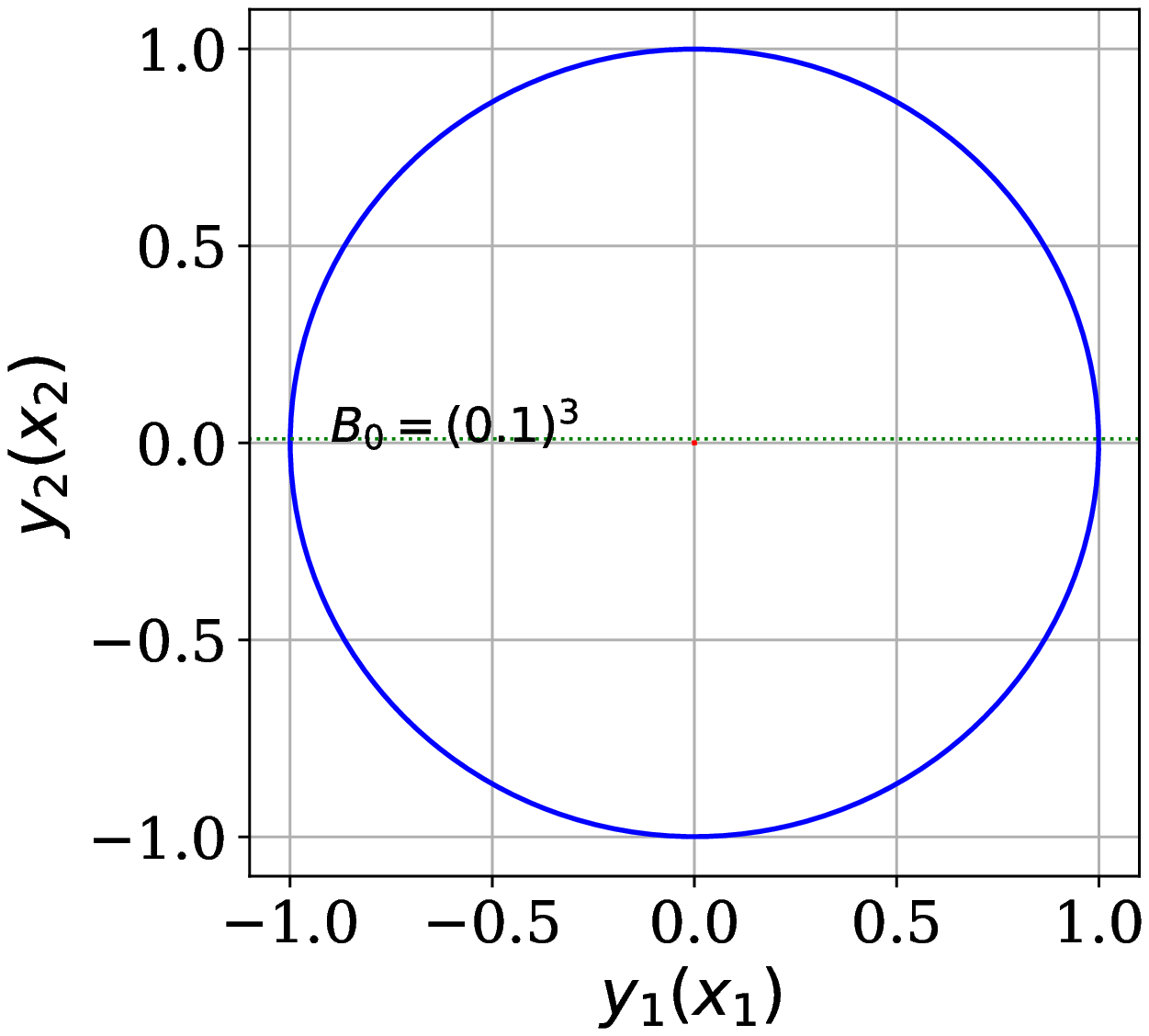}
\includegraphics[width=5.7cm]{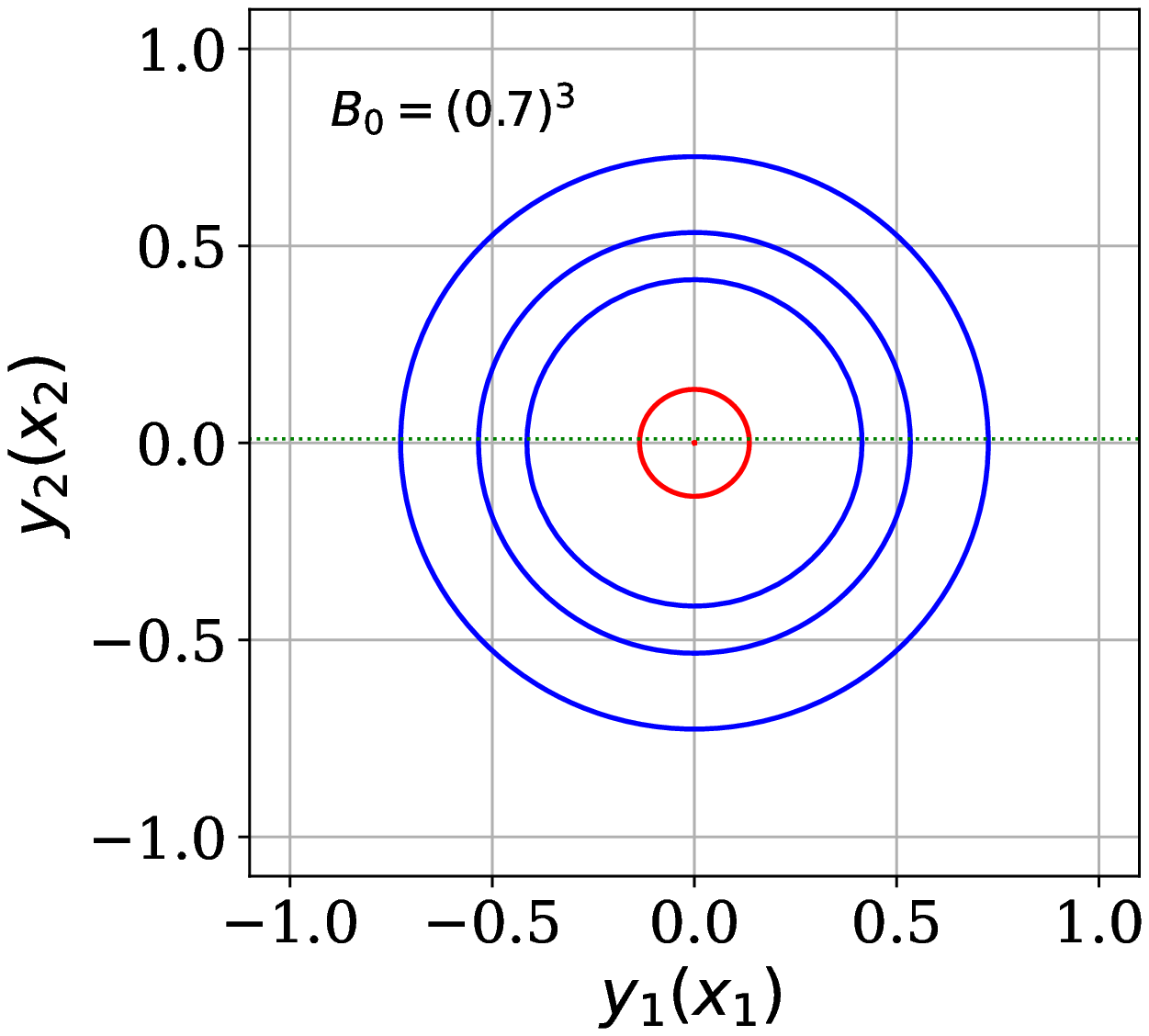}
\includegraphics[width=5.7cm]{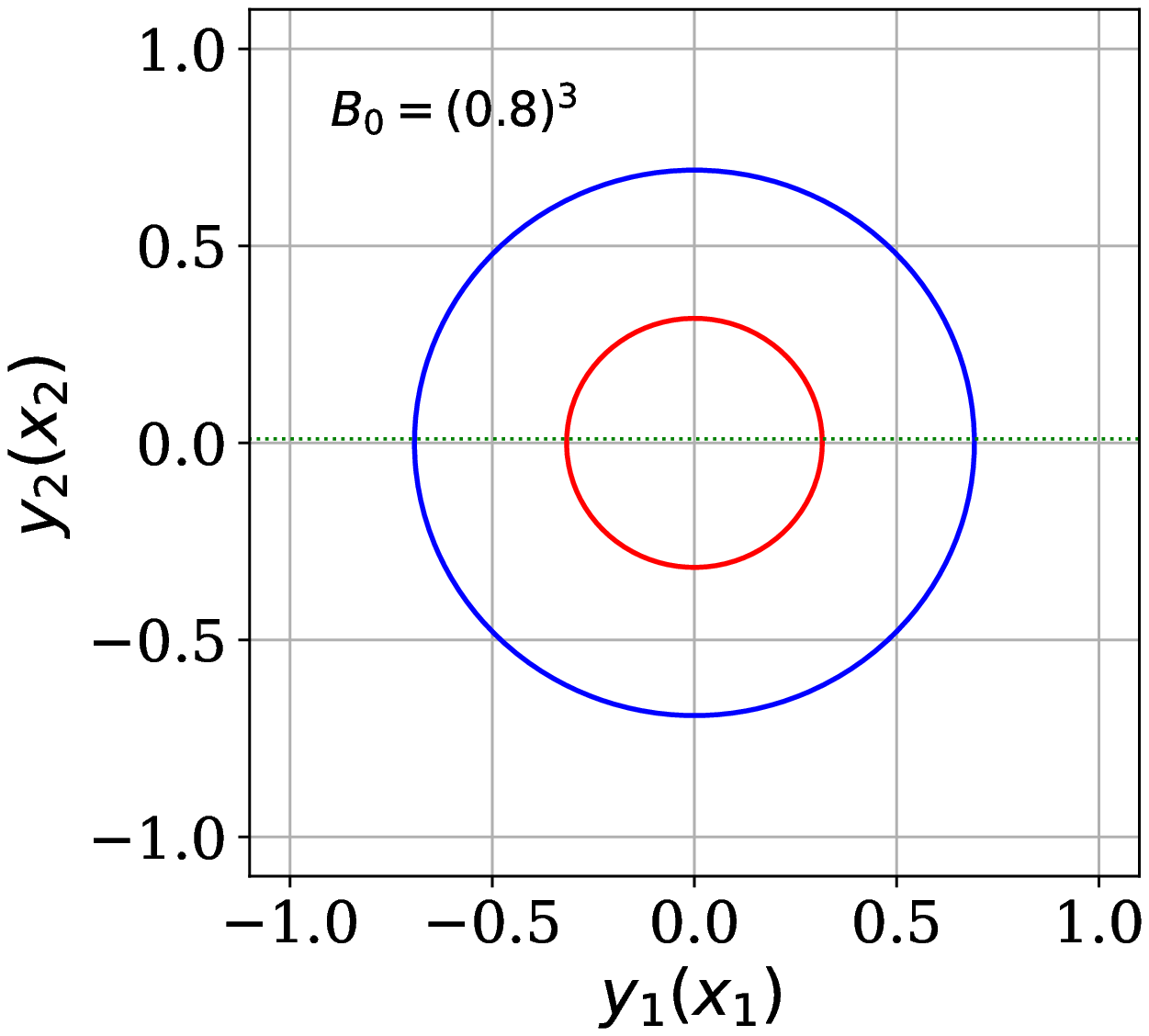}}
{\includegraphics[width=5.7cm]{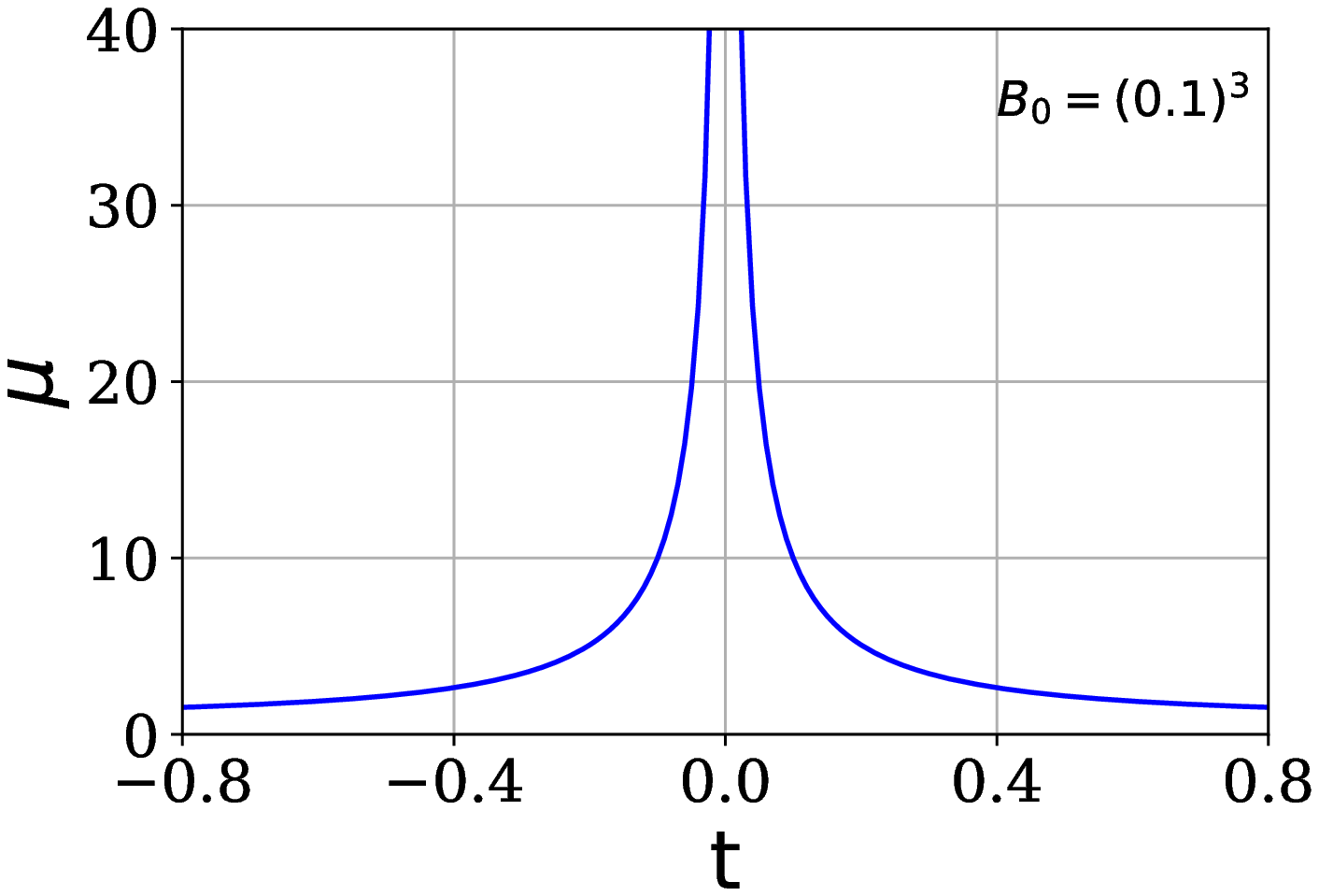}
\includegraphics[width=5.7cm]{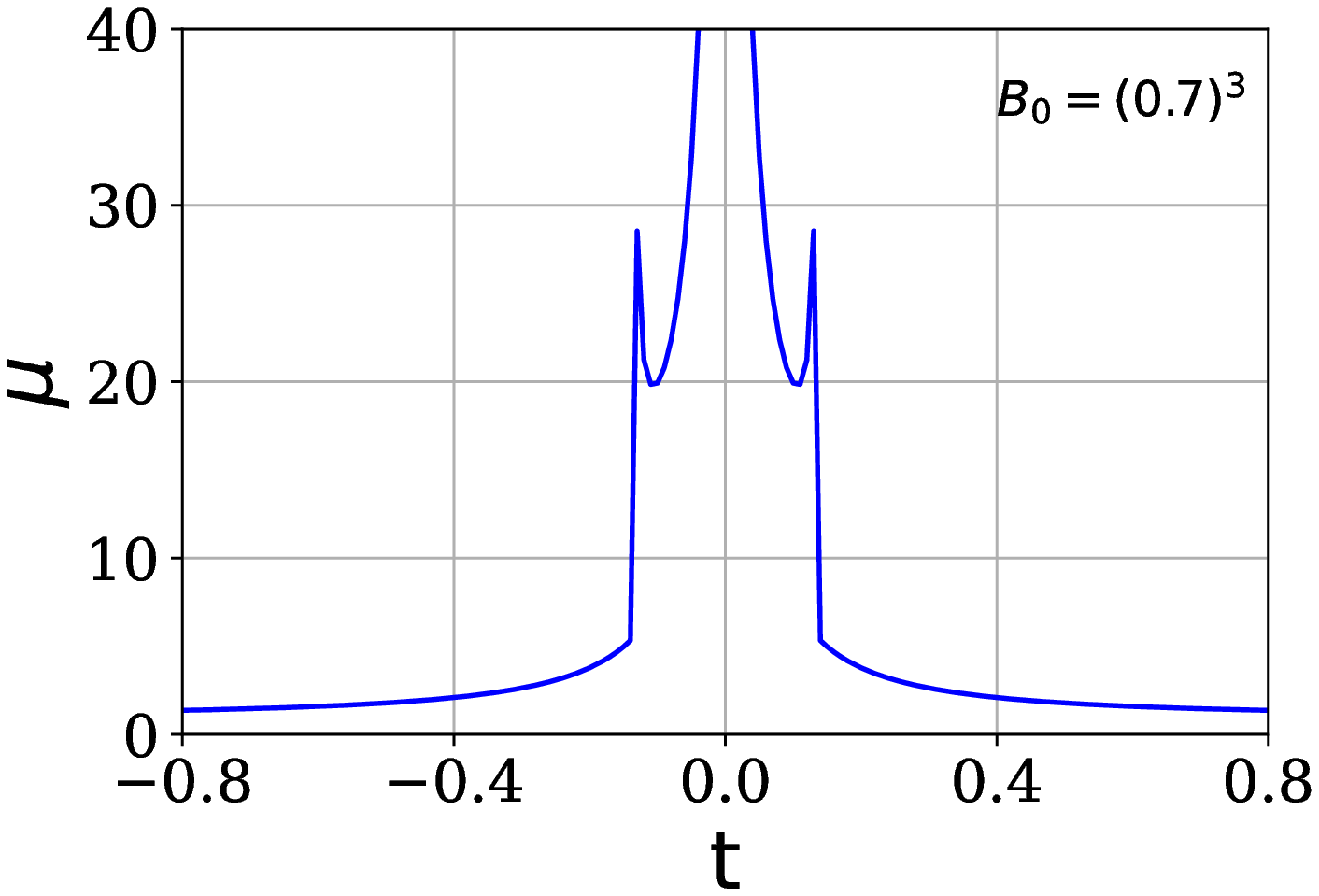}
\includegraphics[width=5.7cm]{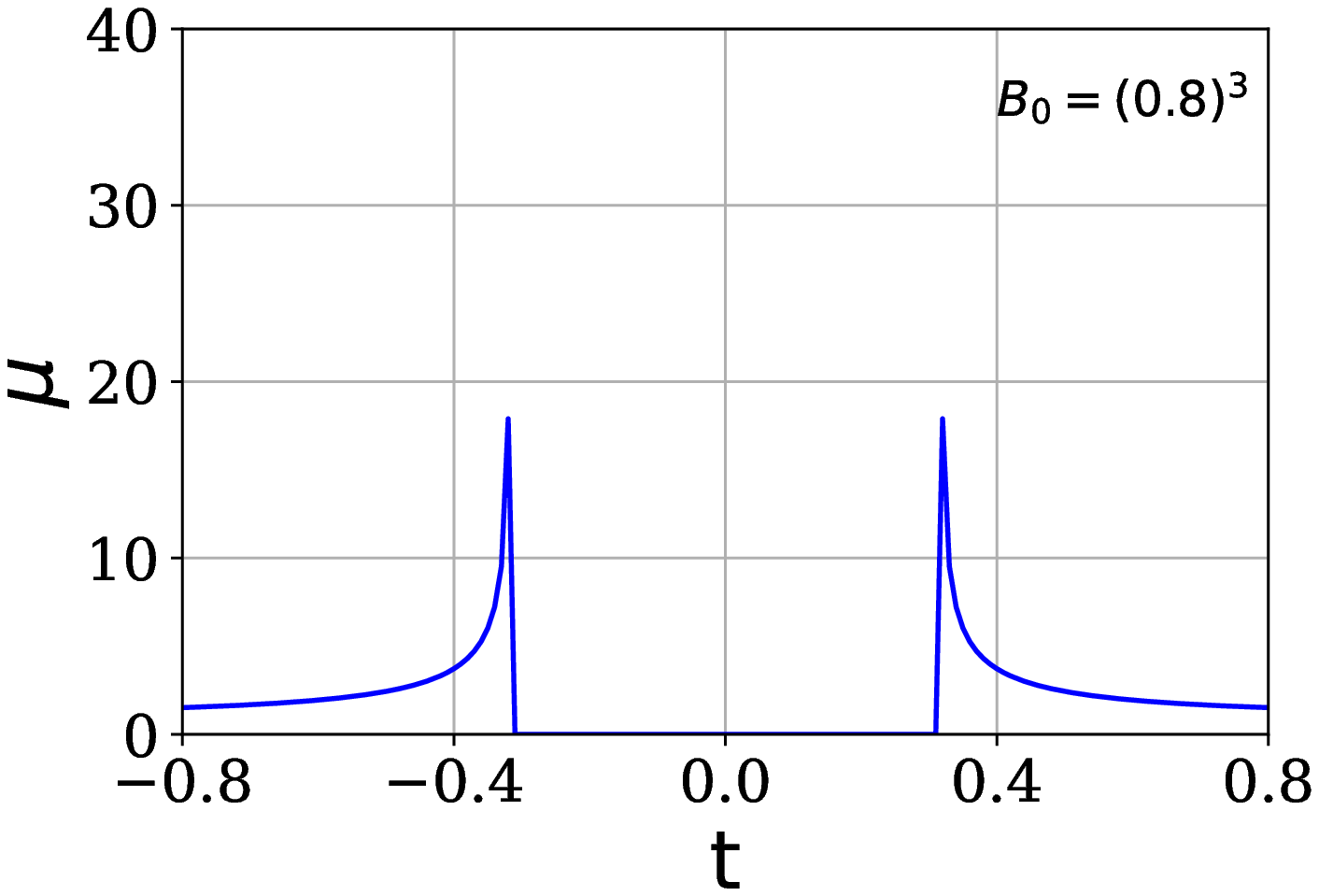}}
\caption{Hill-hole effect on the example of a single lens surrounded by the plasma with power law index $h=2$ (cf. with \citet{2020MNRAS.491.5636T} where $h=3/2,3$ were considered). The upper panels are the critical curves (blue) and caustics (red), the dotted green line is the trace of the background source ($y_2=0.01$); the lower panels are the corresponding magnification curves. In the case when the plasma effect is small (left panels), the microlensing curve is very similar to the vacuum one. In the centre of the source plane there is a caustic in the form of the point. When the plasma effect (given by coefficient $B_0$) increases (central panels), there appears another caustic in the form of a circle and corresponding peaks on the graph; the shape of the 'trident' arises. With further increase of $B_0$, the circular caustic will decrease until it collapses to zero; its corresponding peaks move toward the centre of the microlensing curve. After reaching zero, the caustics in the centre disappear, and there is only one circular caustic that expands as the coefficient $B_0$ increases. A minimum ('hole') appears in the centre of the microlensing curve (right panels).} 
\label{fig:singlestar_cc}
\end{figure*}

In all our calculations, we use $n_0=5.6$ cm$^{-3}$, see Tab.\,\ref{tab:lens_para}. We also note that the electron density near the stars can have large uncertainties. Higher density can be found in the atmosphere of compact objects. It leads to the number density of order of $n_e\sim 10^{19}$ cm$^{-3}$ on the surface of compact objects \citep[e.g. around black hole or neutron star,][]{2016MNRAS.458.3420W,rogers17a}. In such case, one expects to see significant plasma effects in the sub-mm bands, or even in the infrared bands. 
Moreover, higher electron density can be found in accretion disc around a black hole \citep[e.g.][]{2006ApJ...637..968A}, about $10^{-5}$ g/cm$^3$, see also \citet{2011stph.book.....B}. One thus expects some complicated microlensing effects in the strong lensed QSOs.

In order to obtain the magnification curves, we move a source behind the the lens system, and calculate the magnification caused by the lens system. It is possible that the lens can generate multiple images. However, due to the small image separations, we assume that the multiple images cannot be resolved spatially, and thus calculate the total magnification. In most cases, the magnification will be dominated by only one image. We follow the similar approach in \citet{2020MNRAS.491.5636T} to solve the lens equation and calculate the magnification for each source position. More details about the method of calculation can be found in the Appendix \ref{sec:app-A}.

We begin with the simple case of a single star surrounded by plasma (i.e. no planet, $q=0$). We consider this case for comparison with the work in \citet{2020MNRAS.491.5636T}, and also as an illustration to explain the 'hill-hole' effect. The magnification curve and caustics for the power-law plasma density profile with $h=2$ are shown in Fig.\,\ref{fig:singlestar_cc}. With different values of $B_0$, one can obtain regular single peak, trident, or volcano hole shapes of the curves.

\begin{figure}
\centerline{\includegraphics[width=8cm]{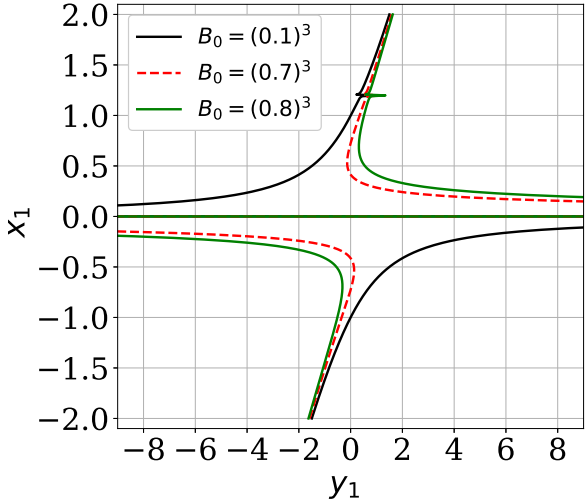}}
\caption{
Lensing mapping (Eq.\,\ref{eq:reduced-lenseq}) for different $B_0$, which corresponding to different frequency of observation. The black, red and green curve presents that at 21 GHz, 1.1 GHz and 0.93 GHz respectively. }
\label{fig:young1}
\end{figure}

\begin{figure*}
\centerline{
\includegraphics[width=6cm]{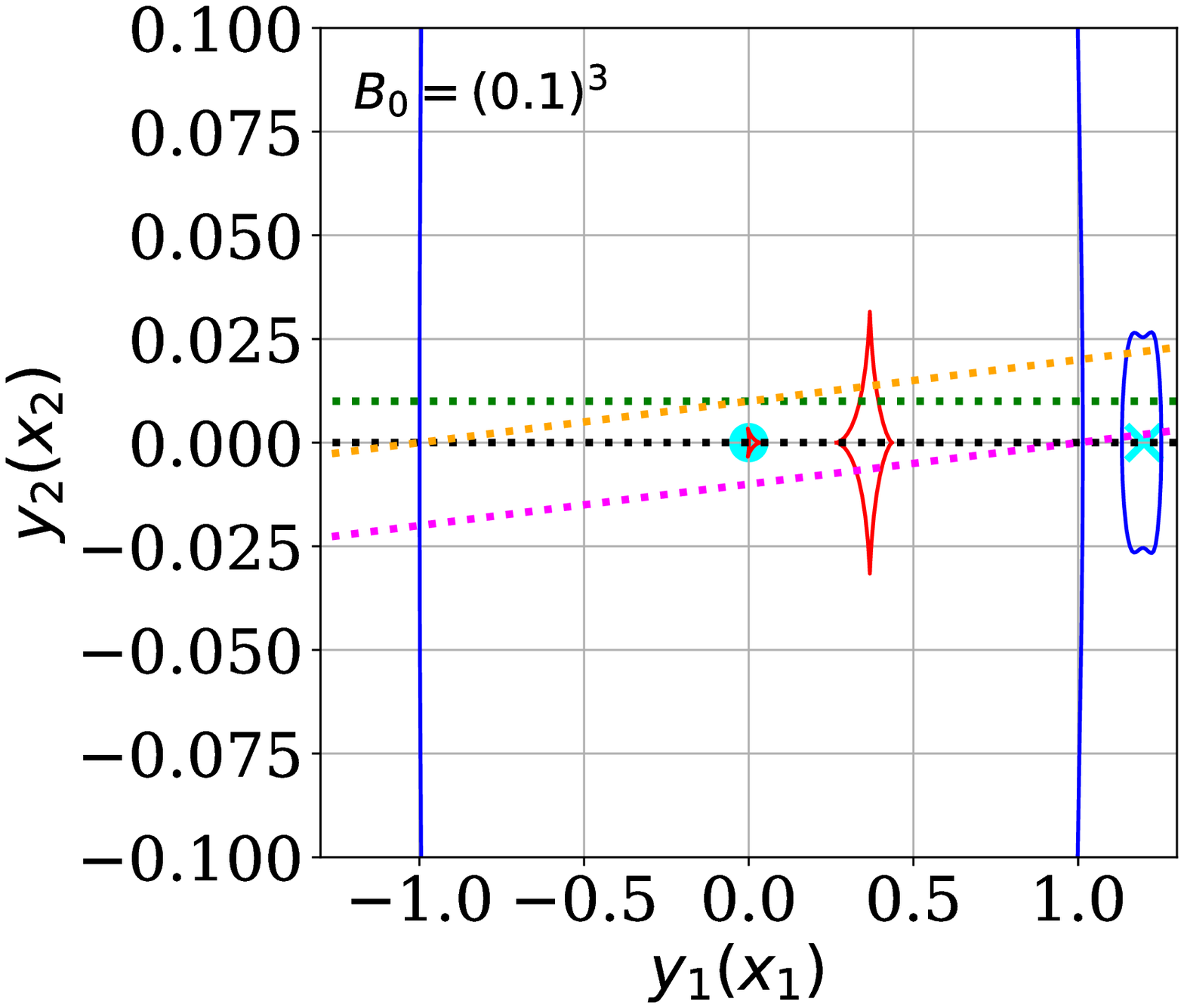}
\includegraphics[width=6cm]{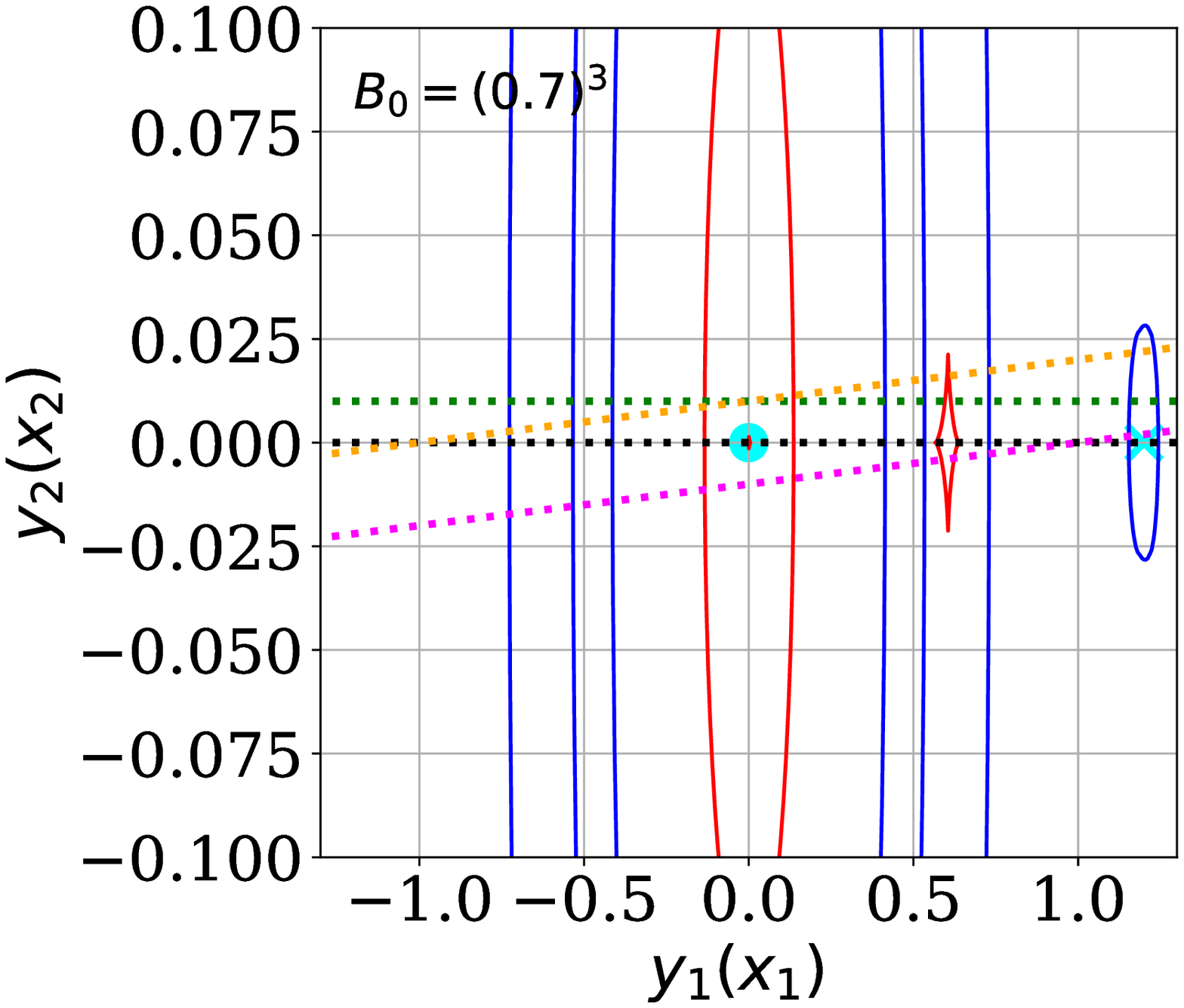}
\includegraphics[width=6cm]{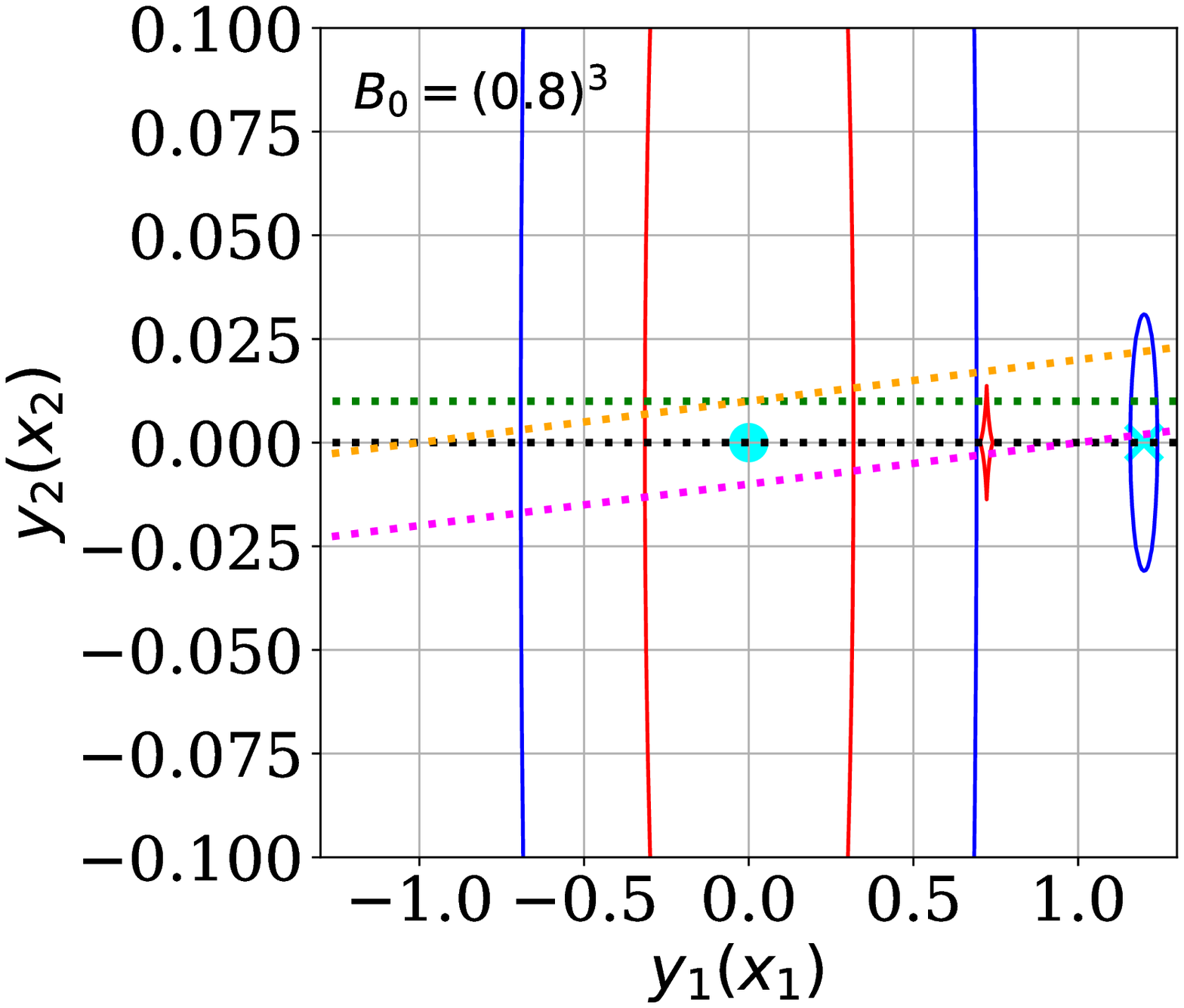}}
\caption{
Lensing structure of binary system for different plasma strength $B_0$ together with the traces of the background source (compare with upper panel of Fig.\,\ref{fig:singlestar_cc} where the single point lens is considered for the same values of $B_0$). The star is at the origin which indicated by the cyan circle, and the cyan cross represents the position of the planet. The blue (red) curve presents the critical curve (caustics) generated by the lensing system. The black, green, orange and magenta lines show the different traces of the background source. From left to right the plasma scale angular radius is $\theta_0=0.1,0.7,0.8 \theta_E$ respectively. All the figures are compressed in $y_1(x_1)-$direction for better visibility. The corresponding microlensing curves are shown in Fig.\ref{fig:mag-combined}.
}
\label{fig:trace}
\end{figure*}

\begin{figure*}
\includegraphics[width=8.5cm]{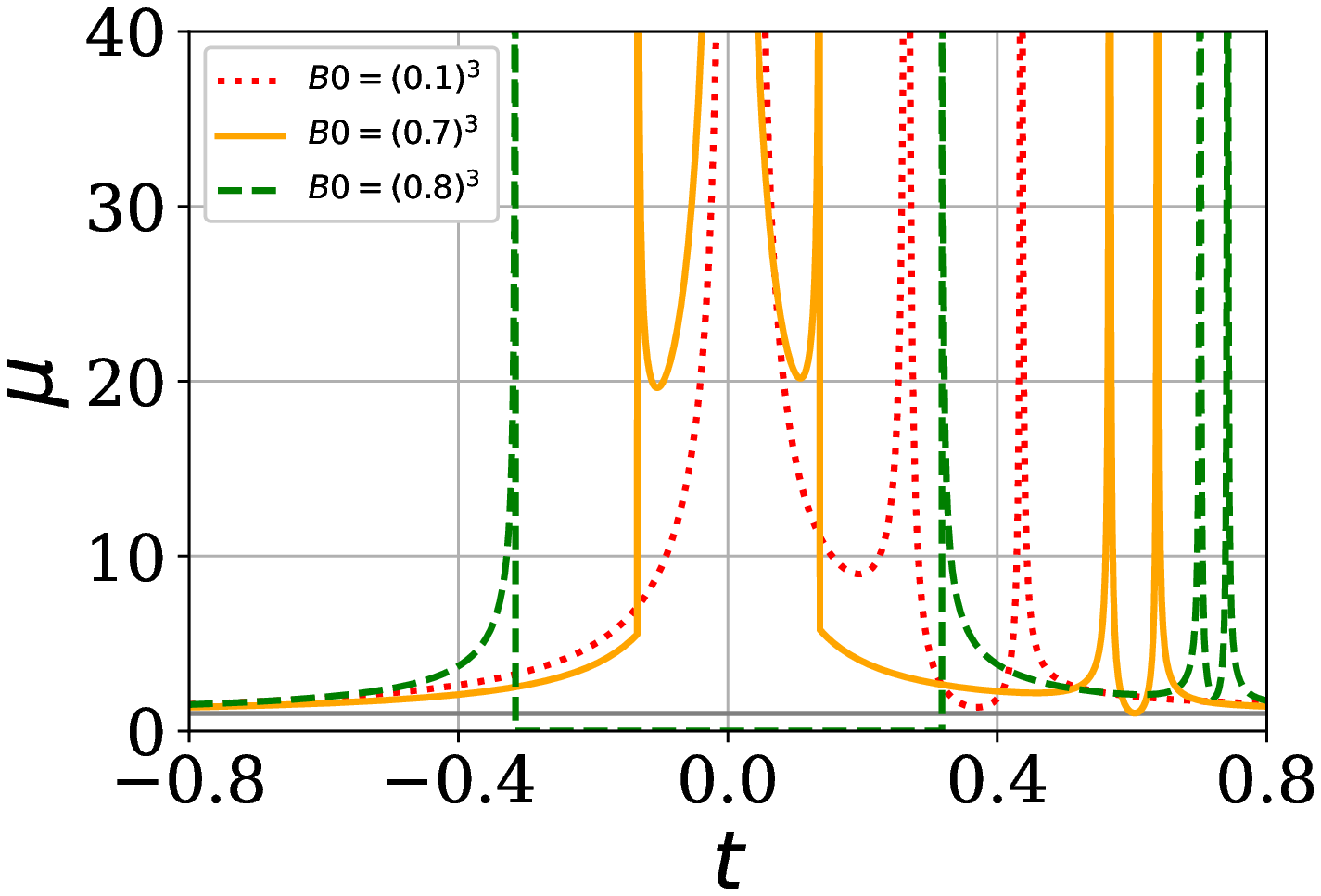}{1}
\includegraphics[width=8.5cm]{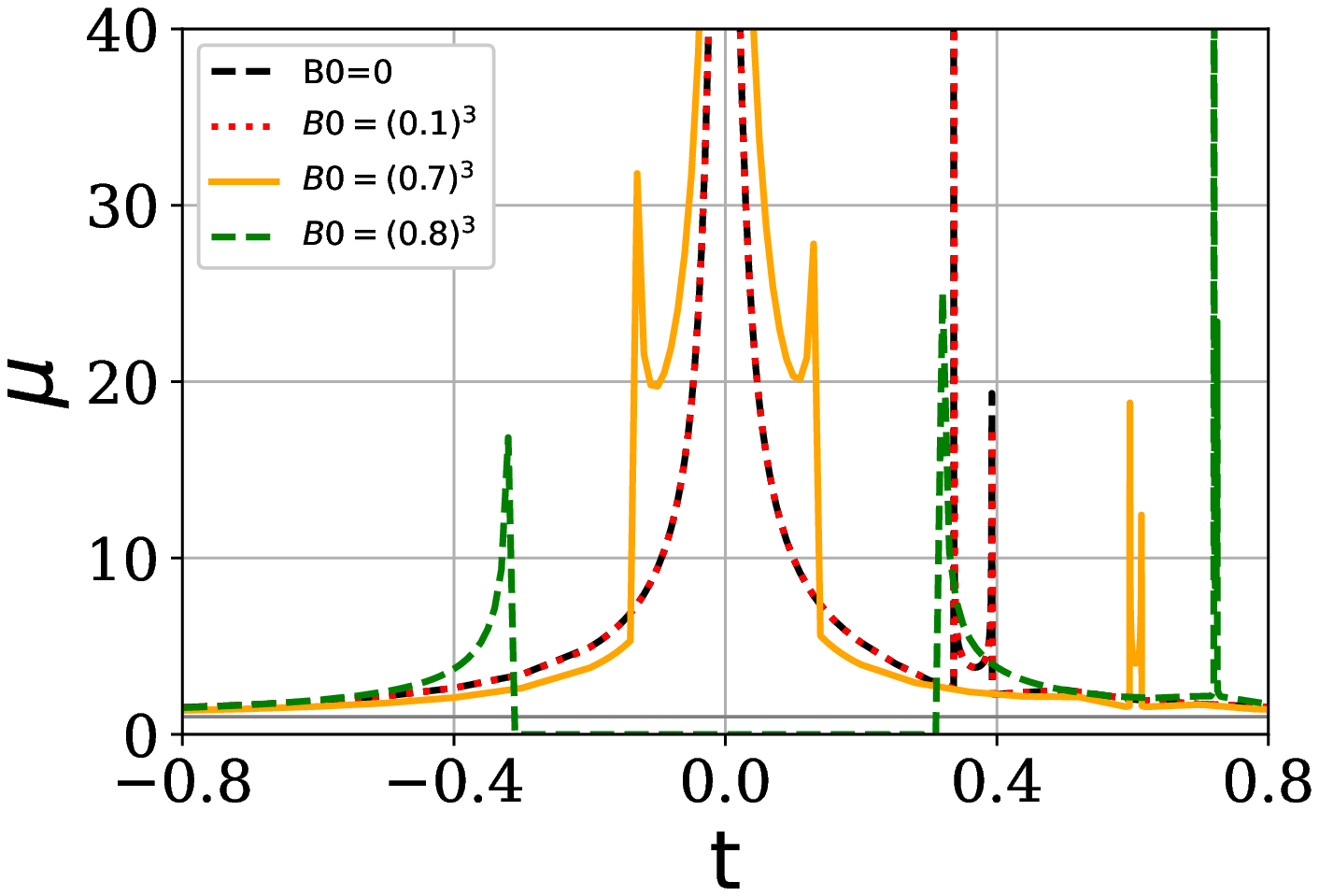}{2}
\\
\includegraphics[width=8.5cm]{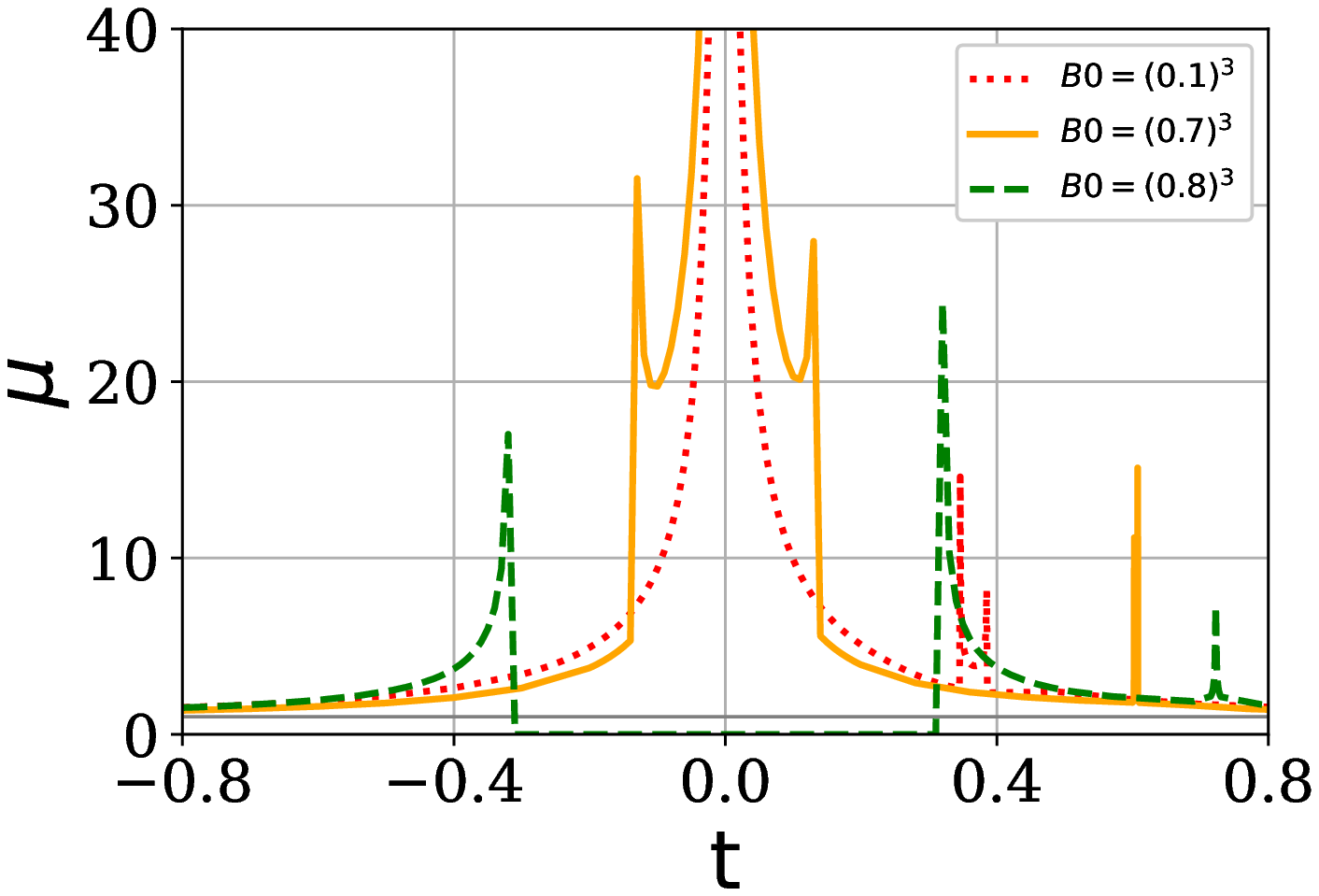}{3}
\includegraphics[width=8.5cm]{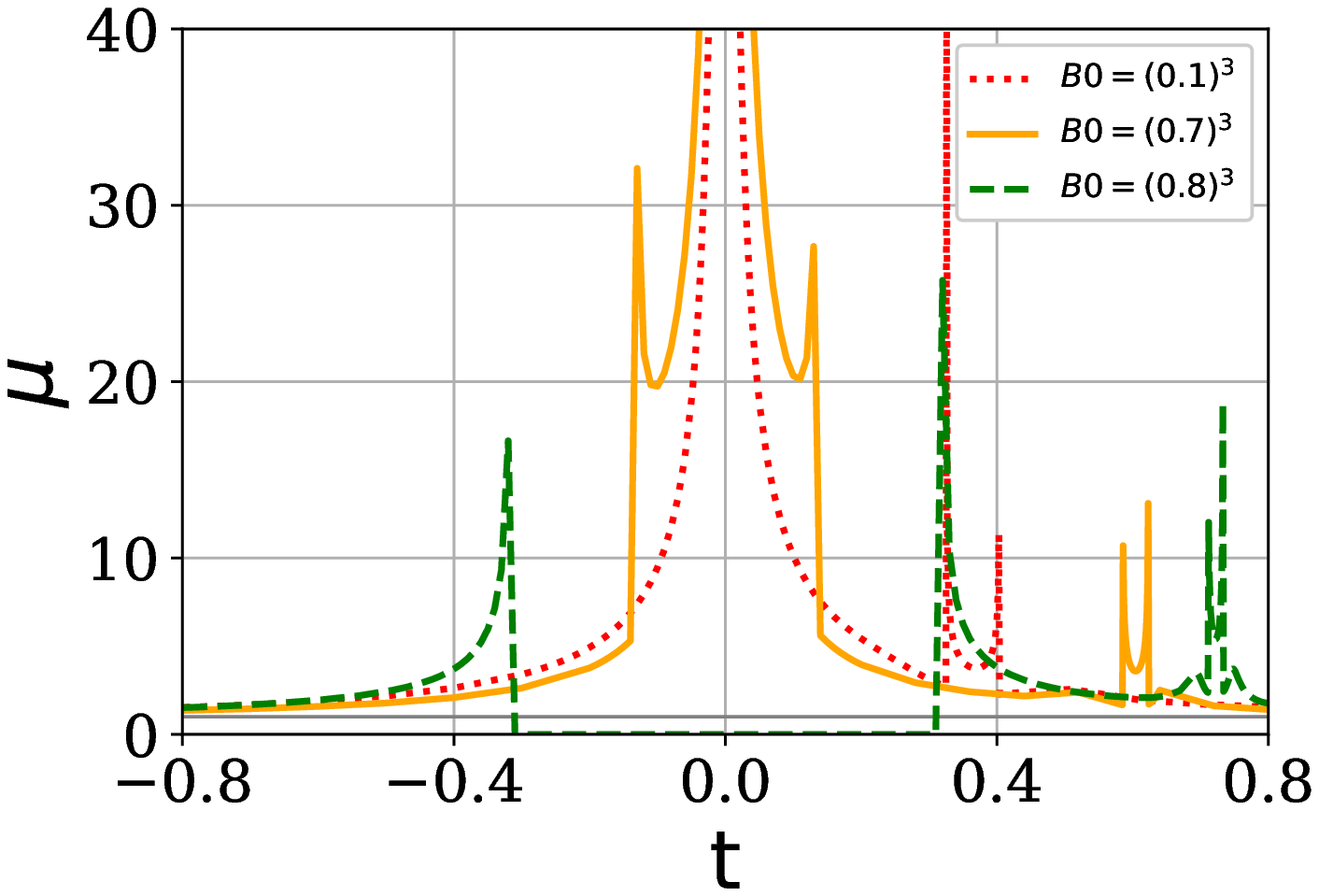}{4}
\caption{
Magnification curves with different $B_0$ and different traces shown in Fig.\ref{fig:trace}. Each panel corresponds to one of four traces from Fig.\ref{fig:trace} with three different values of $B_0$.
Panel (1) corresponds to the black trace in Fig.\,\ref{fig:trace}, panel (2) corresponds to the green trace, panel (3) corresponds to the orange trace, panel (4) corresponds to the magenta trace. For panel (2) the vacuum case ($B_0=0$, black dashed curve) is also given; it is almost indistinguishable from the plasma case with small $B_0$ (red dashed curve). }
\label{fig:mag-combined}
\end{figure*}

With the inclusion of a planet ($q=0.001$, $\vc{x_p}=(1.2,0)$), the magnification curves will be perturbed, and will be more complicated. In the figure of lensing map along $x_1(y_1)-$axis (Fig.\,\ref{fig:young1}), we can see that when the plasma density is low (or the observational frequency is high), the plasma deflection is weak, and similar to the case in vacuum.
The lensed images caused by the star will be on the both sides of the lens, while the perturbations caused by the planet will always on the same side of the planet, even when the deflection caused by plasma becomes strong. In Fig.\,\ref{fig:trace}, we show the critical curves (blue) and caustics (red) for all three cases of $B_0$ that are considered for single lens on Fig.\ref{fig:singlestar_cc}. In the left panel, they are almost identical to the case in vacuum. With the increasing of $B_0$ the plasma firstly generates another caustics. They also pushes the caustics of planet outwards, and compress the caustic generated by the planet.

Several impact parameters and inclination angles of the trace of the background source are studied, as shown by the dotted straight lines in Fig.\,\ref{fig:trace}. The black trace connects the star and the planet, and the green trace is parallel to the black one with an offset of $y_2=0.01$. The other two have an inclination to the axis of star and planet. The functions of the four traces are given by
\begin{align}
y_2&=0, \quad y_1=t, \qquad {\rm black};
\label{eq:trace1} \\
y_2&=0.01, \quad y_1=t, \qquad {\rm green};
\label{eq:trace2} \\
y_2&=0.01 y_1+0.01, \quad y_1=\cos(\arctan(0.01))t, \qquad {\rm orange};
\label{eq:trace3} \\
y_2&=0.01 y_1-0.01, \quad y_1=\cos(\arctan(0.01))t, \qquad {\rm magenta},
\label{eq:trace4}
\end{align}
where we write the trajectory of the source as a function of time $t$. The time $t$ is in unit of $t_E$ given an arbitrary tangential velocity of the source in the source plane. In the choice of our trace functions, the sources have the same tangential velocity. Then we can calculate the magnification curves $\mu(t)$ for the four traces, by inserting Eqs.(\ref{eq:trace1}) -- (\ref{eq:trace4}) into Eq.(\ref{eq:reduced-lenseq}).  
The total magnification is calculated and shown in Figs.\,\ref{fig:mag-combined}.

In the high observed frequency case (small $B_0$), the magnification curve is almost identical to that in vacuum. One can see that from the black and red curves in Fig.\,\ref{fig:mag-combined} panel 2. The central peak generated by the star behaviours like that without the planet. The plasma deflection weakens the peaks in the magnification curves.  With the increasing of $B_0$ the single central peak becomes trident, and then volcano hole. 
The caustic generated by the planet shrinks. 
The double peaks generated by the planet move outwards of the star. In the intermediate case ($B_0=(0.7)^3$), both plasma and gravity cause significant deflection. Thus we can see all the peaks. Such signatures are distinguished from that without plasma or single star with plasma environment.


\section{The perturbation of caustics}
\label{sec:caustics}

In this Section, we discuss the structure of the caustics of binary system in the presence of plasma and their evolution with changes in parameters.

\begin{figure}
\centering
\includegraphics[width=8cm]{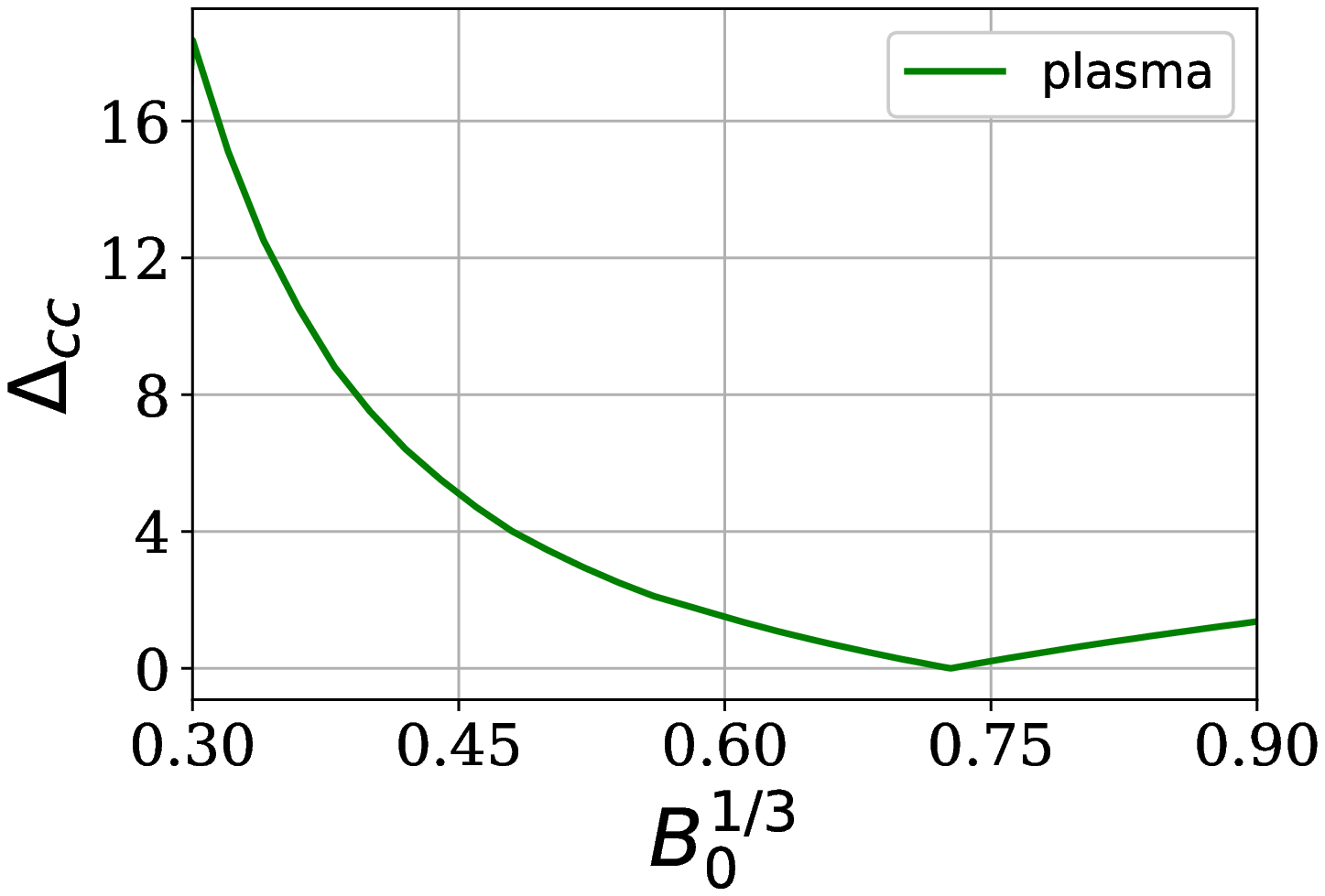}
\caption{
The evolving of the size of plasma caustic as a function of $B_0$ in the case that only a single star surrounded by plasma.
Central caustic ('gravitational one') is a point, and it disappears after merging with plasma caustic in the origin of coordinates.
}
\label{fig:sep_withoutplanet}
\end{figure}

First, let us comment on the caustics for the case of a single lens surrounded by plasma (Fig.\,\ref{fig:singlestar_cc}). For $h=2$ power law profile, there are two caustics: central as a point (associated with gravity) and a circular (associated with plasma). As long as the plasma influence is small, the plasma caustic has a large radius. E.g., in Fig.\,\ref{fig:singlestar_cc} for $B_0=0.1^3$, the radius of the plasma caustics extends beyond the figure. For central panels in Fig.\,\ref{fig:singlestar_cc}, where the parameter $B_0$ is larger ($0.7^3$), we see both caustics: the red dot in the centre of the figure and the red circle around it. With the further increasing of the $B_0$, there will be an transition, i.e., the hole in the magnification curve forms and the central caustic disappears at the same time. 
We use a horizontal width $\Delta_{cc}$ to indicate the size of the caustics, which is the diameter of the caustics in an axis-symmetric case. In Fig.\, \ref{fig:sep_withoutplanet}, we show the evolving of the size of caustics as a function of $B_0$. Since the central caustic is a point, its $\Delta_{cc}$ is zero. With the plasma density increasing further, the plasma caustic shrinks and finally merges with the central caustic. One can see that there is only one (plasma) caustic from the right panel of Fig.\,\ref{fig:singlestar_cc}. After merging, the caustic will increase with the influence of the plasma, $B_0$.  
The transition without planet has been given analytically \citep{2020MNRAS.491.5636T}, see Appendix \ref{sec:hole-formation} for more detail. The critical value is $B_0=2/3^{3/2}\approx 0.3849$ ($B_0^{1/3}\approx 0.7274$, and the corresponding frequency in our density model is around 1.07 GHz).

Now let us consider a binary system where the star is surrounded by plasma. Note that even for the vacuum case, the existence of the planet will generate extra caustics and change the central caustic as well \citep[e.g.][]{1999A&A...348..311B}. It has been approximately studied in an analytic way or numerically in vacuum \citep[e.g.][]{2005MNRAS.356.1409A,2005ApJ...630..535C,2021arXiv210207950A}. However, with the extra deflection by the plasma, the analytic form is difficult to obtain.

\begin{figure*}
\includegraphics[width=8cm]{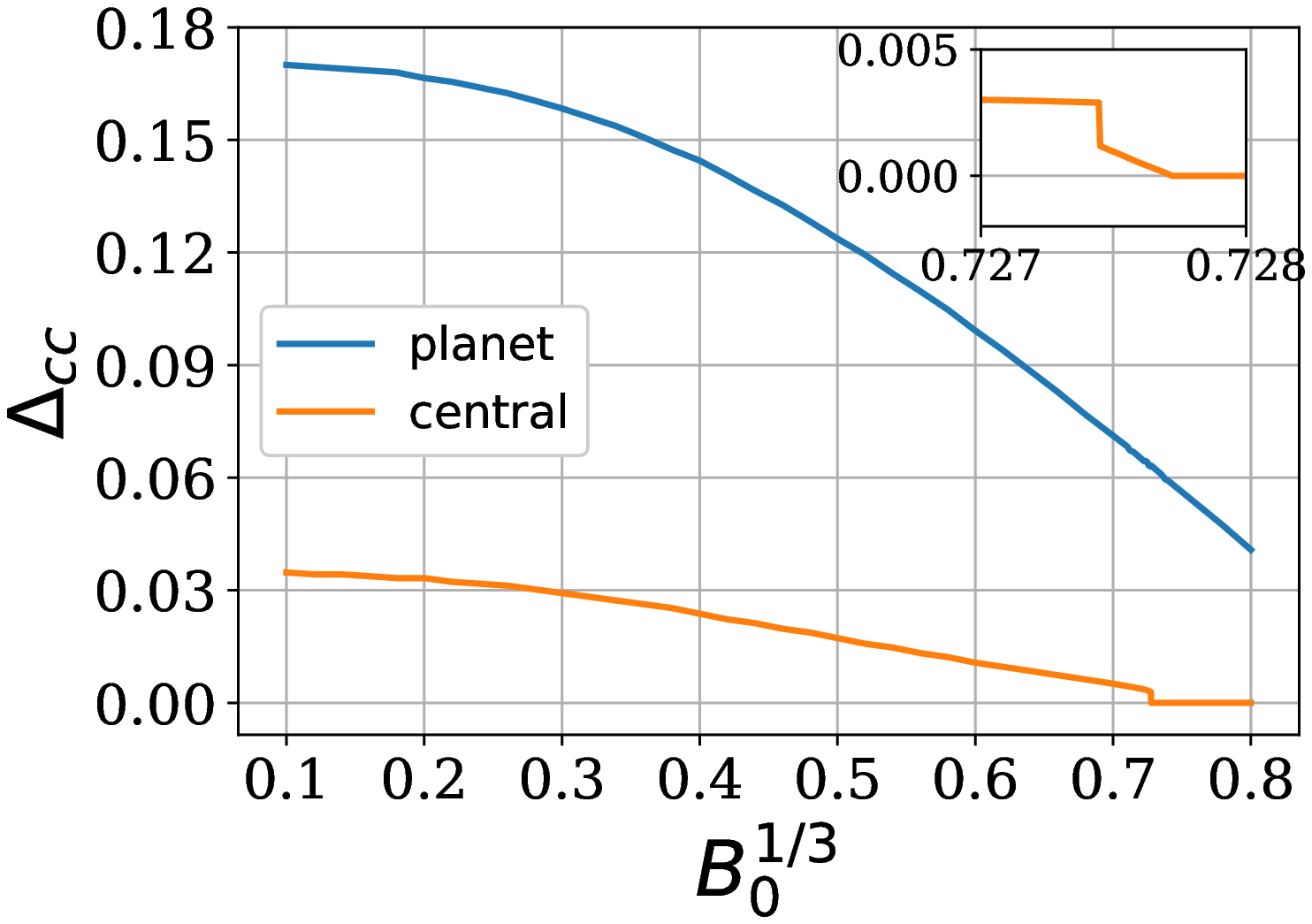}
\includegraphics[width=8.25cm]{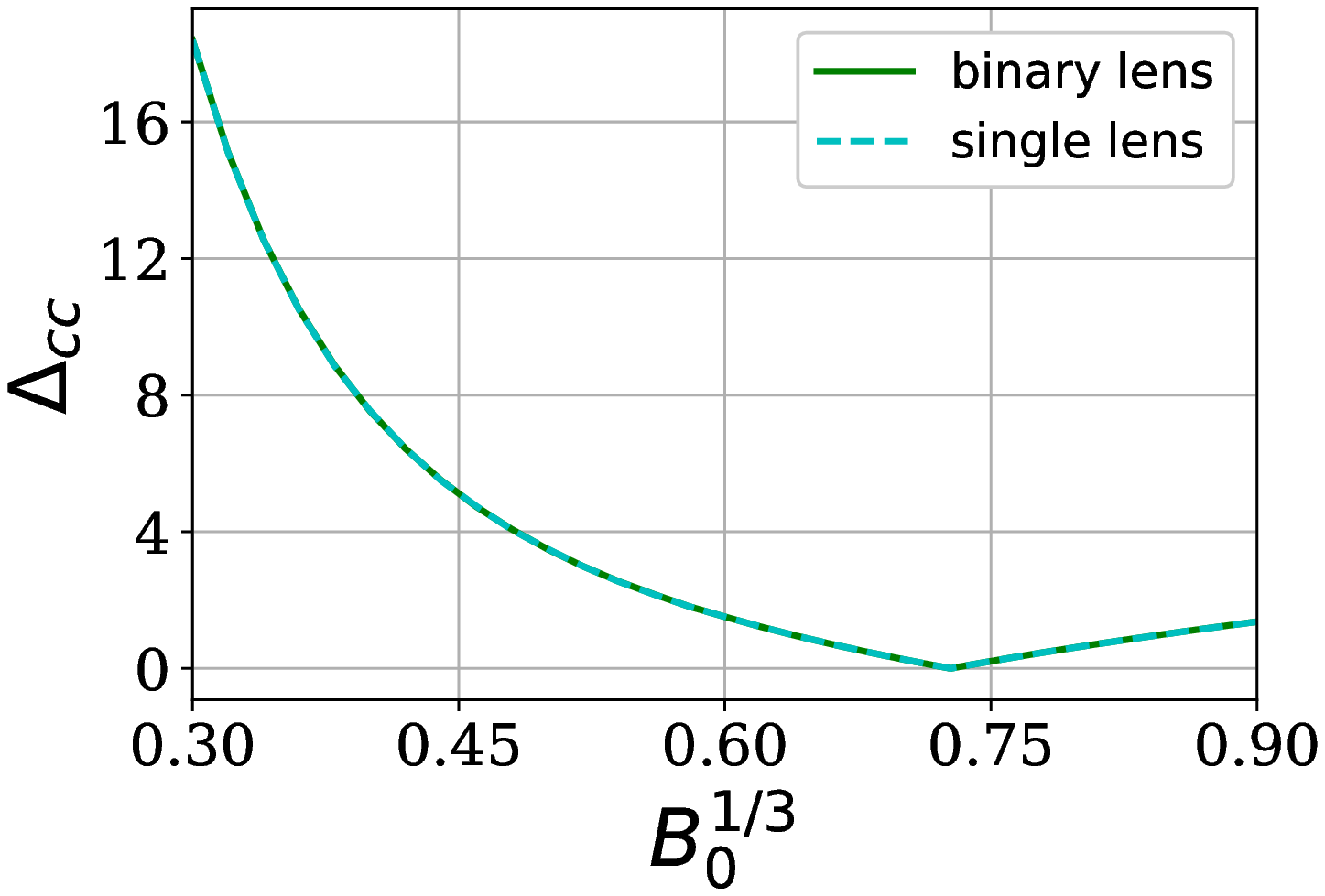}
\caption{
The left picture shows the size of the central caustics (orange) and planetary caustic (blue) of a binary system in which the star is surrounded by plasma (In case there are multiple central caustics, we focus on the largest caustic), the green solid line in the right picture shows the size of the caustic caused by plasma, all of them measured by the horizontal width, as a function of $B_0$ ($x_p=1.2$). The image in the upper right of left panel is an enlarged portion of the size of the central caustics. The transition of the caustics (the merge of part of the central caustics and the plasma caustic) around $B_0^{1/3} \sim 0.7275$ is shown in Fig.\,\ref{fig:app-caustics7278}. 
The right panel is almost identical to Fig.\,\ref{fig:sep_withoutplanet} (shown as the dashed line in the right picture), but the other lensing effects can show small difference, e.g. the positions of peaks in magnification curve.
}
\label{fig:dcc-withB0}
\end{figure*}

\begin{figure*}
\includegraphics[width=8cm]{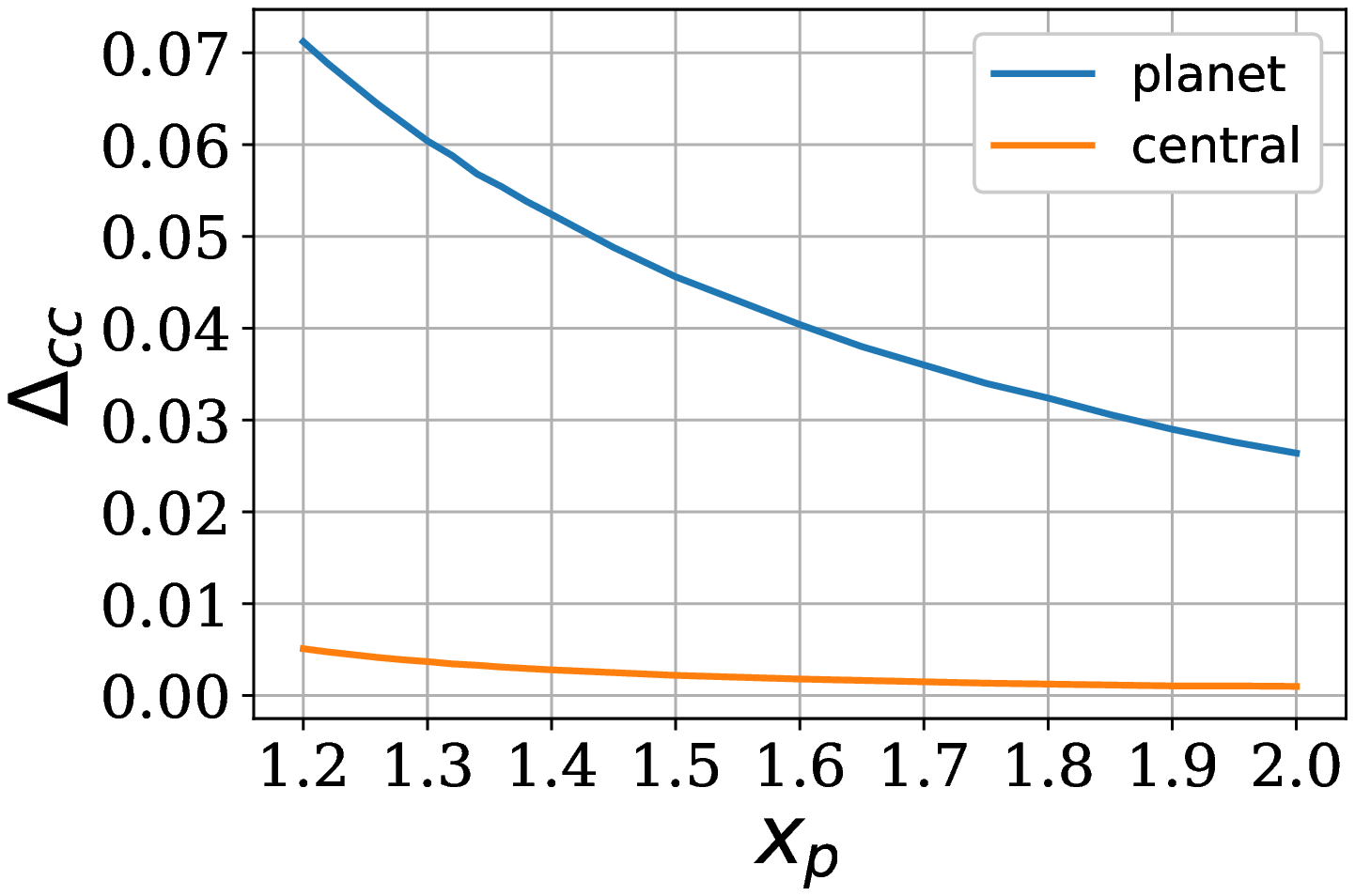}
\includegraphics[width=8.2cm]{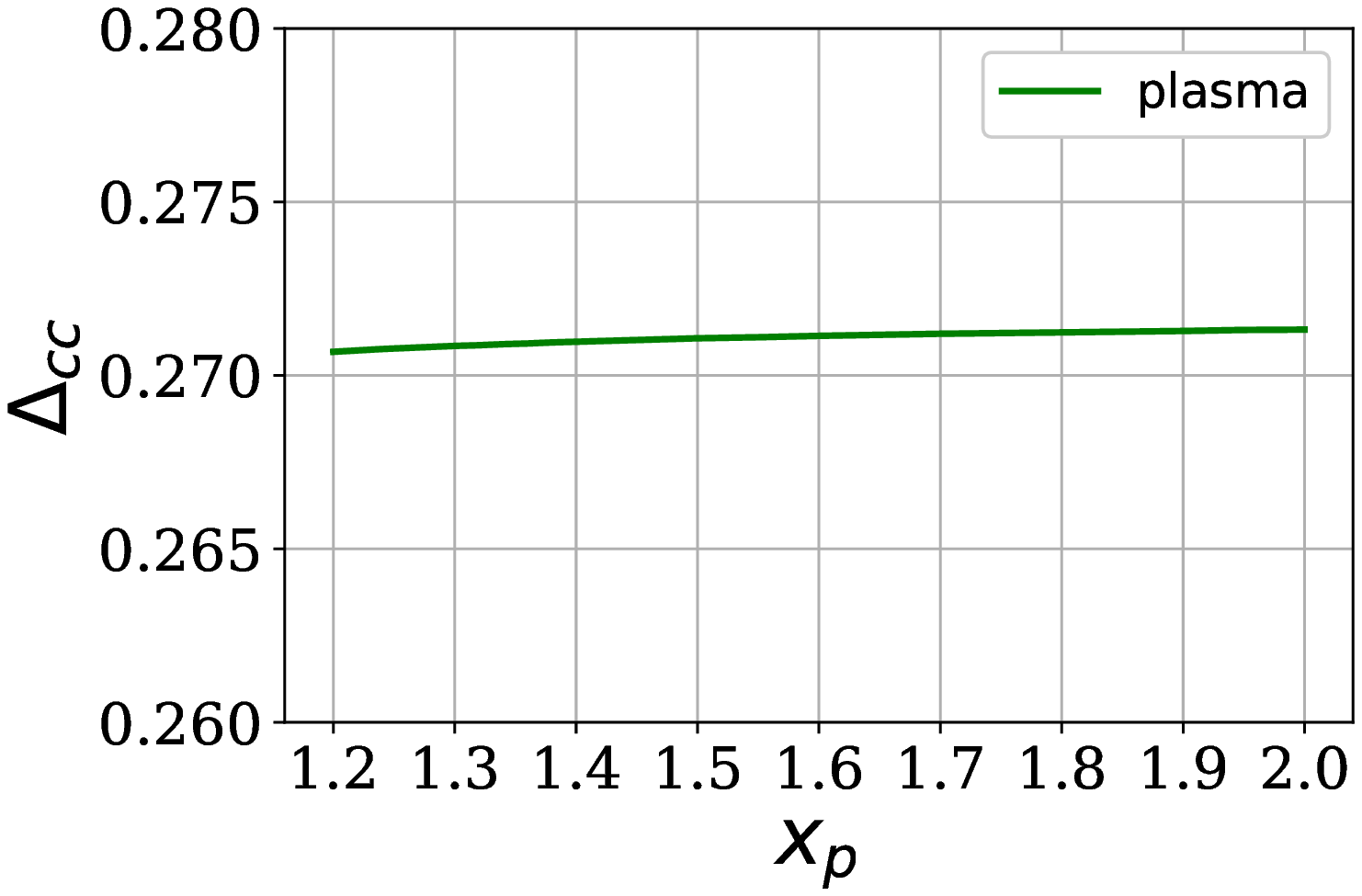}
\caption{Same as Fig.\,\ref{fig:dcc-withB0}, but as a function of $x_p$, and $B_0=(0.7)^3$.}
\label{fig:dcc-withxp7}
\end{figure*}

In presence of plasma in binary system, more caustics will be generated by the lens, e.g. more than one central caustics, the one associated with the planetary and the one associated with plasma. Here we only consider a simple case: a large separation between the star and planet, the two caustics generated by star and planet do not merge together. One interesting difference exists from the case of single star with plasma environment. In case of single star, when the plasma becomes sufficient strong, all the light rays will be diverged, i.e. no solution of image at all. However, when there is a planet, the gravity of the planet will weaken the diverging caused by the plasma. One solution of image to the lens equation exists, although the magnification of this image can be extremely small. Only when the refraction deflection by plasma is strong enough ($B_0^{1/3} \sim 5$), there will be no solution at all. Thus the definition of hole in the magnification curve is different from that in the single star case. We consider the hole forming when there is only one image with small magnification, $< 10^{-5}$. In our lensing system, it will happen when $B_0^{1/3}\approx 0.7273$ (Appendix \ref{sec:hole-formation}). We can estimate the number of caustics by two ways: first we can use the number and the positions of peaks in the magnification curve; on the other hand, we can see from the pattern of the caustics. We find that the peak number drops from six to four when $B_0^{1/3}\approx 0.7278$, two peaks related to the central caustics disappear, which imply that the central caustics disappear. 
The caustics can have either diamond-like shapes (four cusps) or triangle-like shapes (three cusps) \citep[e.g.][]{1999A&A...348..311B,2022arXiv220712412Z}. One or two of the cusps are located on the star-planet axis, i.e. $y_2=0$. We thus define the width $\Delta_{cc}$ as the distance between the points where caustic crosses an axis connecting the star and planet. In Figs.\,\ref{fig:dcc-withB0}-\ref{fig:dcc-withxp7}, we show the evolving of $\Delta_{cc}$ with $B_0^{1/3}$ and $x_p$. 
From left panel of Fig.\,\ref{fig:dcc-withB0} we can see that both caustics shrink with the increasing of the plasma effects. While when $B_0^{1/3}>0.93$, the size of planetary caustic will increase again with $B_0$ (we do not consider a case with such low frequency of observation, thus not show in the figure).
When $B_0^{1/3}\sim 0.7275$, the caustics caused by plasma and part of central caustics merge (green curves in Fig.\,\ref{fig:app-caustics7278}). The other caustic which is not merge with the plasma caustic is regarded as the central one. We can see the transition, i.e. a drastic decreasing of the size of the central caustics. The size of the caustic generated by plasma have almost the identical behaviour as that without planet (Fig.\,\ref{fig:sep_withoutplanet}). Only the  transition point is slightly different, $B_0^{1/3}\sim0.7275$. 
The caustic generated by the planet is larger than that by the central one. 
In all cases, $\Delta_{cc}$ of planet decreases with $B_0^{1/3}$ and $x_p$. The position of the planet caustics will be changed by plasma as well, but no transition has been found.

In Fig.\,\ref{fig:dcc-withxp7} (left), we present the evolving of the size of caustics as a function of the star-planet separation $x_p$. As one would expect, the larger the separation the smaller the caustics. Additional tests with different $B_0$ have been performed, and similar trend has been found. For the size of the caustics generated by plasma, the planet has little effect. As shown in Fig.\,\ref{fig:dcc-withxp7} (right), $\Delta_{cc}$ increases with $x_p$ slowly, for the dependence on $x_p$ of plasma caustic is much weaker compared with that of other caustics on $x_p$.

\section{Summary and Discussion}
Microlensing provides a powerful tool in searching and studying the exoplanets. Taking into account of the stellar wind (plasma), the magnification curves generated by the system can be different at low frequencies. This happens due to refraction in the plasma, which leads to chromatic deflection of the light ray in addition to gravitational deflection. We present some magnification curves with different observational frequency and different trajectories of the background source for the system of a star surrounded by plasma and a planet (Figs.\,\ref{fig:trace} and \ref{fig:mag-combined}). When the observational frequency is high, the magnification curves are almost indistinguishable from those without plasma. But as the observational frequency decreases to a few GHz, the plasma effects will be significant. For example, the peak in the magnification curve generated by the star becomes trident or even volcano hole. The position of the peak generated by the planet changes with frequency as well.

We also present the evolving of the size of the caustics with plasma density ($B_0$) and the separation of star and planet ($x_p$) in Figs.\,\ref{fig:dcc-withB0} and \ref{fig:dcc-withxp7}. With the increasing of $B_0$, the size of the central caustics and planetary caustic decrease gradually, and the central caustics disappear after $B_0$ reaching a critical value. For the size of the caustic caused by plasma, it decreases at the beginning and then increases. 
With $x_p$ increases, the size of the central caustics and planetary caustic decrease, but the size of the caustic caused by plasma almost unchanged.

In our study, we find some difference from the case without planet. When the lens is consisted of star and plasma only, the hole forms and the central caustic disappears at the same time, and the caustics have the regular circular shapes. While with the inclusion of a planet, the transition value of the plasma is different. More caustics will be generated, and the shapes of caustics can be diamond-like, triangle-like or even more complex shape. The hole in the magnification curve will form before the central caustics disappear. 
The possible reason is that the existence of the planet slightly shifts the central caustics to the right, thus the disappearance of the central caustics requires a stronger plasma.

The plasma density that we adopted in this study is from the Solar wind. The possible observational frequency is around $\sim$GHz. There is possibility of higher plasma density or density gradient, and thus even in higher frequency, e.g. infrared, we can distinguish and study the stellar wind. 
There are several simplification in this work. First of all, we only study the system of one star and one Jupiter size planet. Binary stars with planet(s) or single star with multiple planets has also been found and will show complicate situations. Even for the system of single star and single planet, there is large parameter space to be explored, e.g. mass ratio, separation etc. 
Both the lens and the background source has been considered as point source, i.e. the finite size of the star or planet has not taken into account. Such effect needs to be included for accurate prediction in the future study. The magnification curve in different bands can provide rich information on the lens system.

\section*{Acknowledgements}
XE would like to thank Chun Xia for the help on the plasma model of the Solar wind. XE is supported by the NSFC Grant No. 11873006, 11933002 and the China Manned Space Project with No.CMS-CSST-2021-A07, No.CMS-CSST-2021-A12. The work of OYuT was partially supported by the Russian Foundation for Basic Research, project No. 20-52-12053. OYuT is thankful to G.S. Bisnovatyi-Kogan for useful discussions.

\section*{Data Availability}
The data underlying this article will be shared on reasonable request to the corresponding author.
\bibliographystyle{mnras}
\bibliography{micropl}

\appendix

\section{Numerical calculation of magnification curves as step-by-step procedure}
\label{sec:app-A}

\begin{figure}
\centerline{\includegraphics[width=8cm]{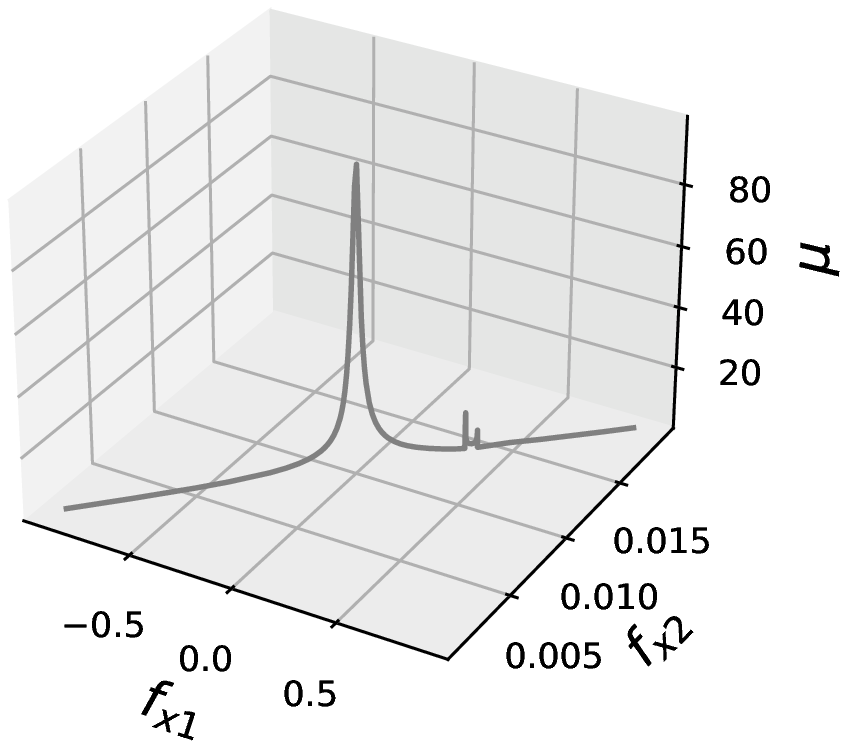}}
\caption{
Magnification curves with $B_0=(0.1)^3$, and the changes of $f_{x1}$ and $f_{x2}$ are consistent with the orange trace of Fig.\,\ref{fig:trace}. }
\label{fig:mag-orange1-3D}
\end{figure}

\begin{itemize}
\item  Following Section \ref{sec:formula}, we can choose parameters $q$, $\vc{x}_p = (x_{p1}, x_{p2})$ and $B_0$, then solve the lens equation. In our case, we use $x_{p1}=x_p$ and $x_{p2}=0$.

\item Choose the trace of the compact source in the source plane by specifying the $y_1$ and $y_2$ as a function of time $t$. In this work we write $y_1 = f_{x1}(t)$ and $y_2 = f_{x2}(y_1)$.
For Fig.\,\ref{fig:mag-combined} (1), $f_{x1}(t)=t,\, f_{x2}=0$. For Fig.\,\ref{fig:mag-combined} (2), $f_{x1}(t)=t,\, f_{x2}=0.01$.
For Fig.\,\ref{fig:mag-combined} (3), $f_{x1}(t)=\cos(\arctan(0.01)) t,\, f_{x2}=0.01 y_1+0.01$.
For Fig.\,\ref{fig:mag-combined} (4), $f_{x1}(t)=\cos(\arctan(0.01)) t,\, f_{x2}=0.01 y_1-0.01$.

\item If we take one of the trace function (e.g. Eq.\,\ref{eq:trace1}) into lens equation (\ref{eq:reduced-lenseq}), it can be written as
$$
\begin{cases}
x_1-\dfrac{x_1}{x_1^2+x_2^2}-q \dfrac{x_1 - x_p}{(x_1 - x_p)^2+x_2^2}+\dfrac{B_0}{x_1^2+x_2^2} \dfrac{x_1}{(x_1^2+x_2^2)^{\frac{1}{2}}} = y_1 \, , \\
\\
x_2-\dfrac{x_2}{x_1^2+x_2^2}-q \dfrac{x_2}{(x_1 - x_p)^2+x_2^2}+\dfrac{B_0}{x_1^2+x_2^2} \dfrac{x_2}{(x_1^2+x_2^2)^{\frac{1}{2}}} = y_2 \, .
\end{cases}
$$
Multiple images can be obtained for some source positions. 
For given source position $(y_1, y_2)$, any numerical solution $(x_1,x_2)$ gives the image. Usually, there are several images.

\item We calculate the magnification of each image
$\mu_i$ with Eq.(\ref{eq:mu}), i.e.
$$
\begin{cases}
\frac{\partial \beta_1}{\partial \theta_1}=1-\frac{\partial \alpha_1}{\partial \theta_1}=1-\frac{d}{d x_1}\eck{\frac{x_1}{x_1^2+x_2^2}+q \frac{x_1-x_p}{(x_1-x_p)^2+x_2^2}-B_0 \frac{x_1}{(x_1^2+x_2^2)^{3/2}}}\\
\frac{\partial \beta_1}{\partial \theta_2}=-\frac{\partial \alpha_1}{\partial \theta_2}=-\frac{d}{d x_2}\eck{\frac{x_1}{x_1^2+x_2^2}+q \frac{x_1-x_p}{(x_1-x_p)^2+x_2^2}-B_0 \frac{x_1}{(x_1^2+x_2^2)^{3/2}}}\\
\frac{\partial \beta_2}{\partial \theta_1}=-\frac{\partial \alpha_2}{\partial \theta_1}=-\frac{d}{d x_1}\eck{\frac{x_2}{x_1^2+x_2^2}+q \frac{x_2}{x_1^2+x_2^2}-B_0 \frac{x_2}{(x_1^2+x_2^2)^{3/2}}}\\
\frac{\partial \beta_2}{\partial \theta_2}=1-\frac{\partial \alpha_2}{\partial \theta_2}=-\frac{d}{d x_2}\eck{\frac{x_2}{x_1^2+x_2^2}+q \frac{x_2}{x_1^2+x_2^2}-B_0 \frac{x_2}{(x_1^2+x_2^2)^{3/2}}} \, .
\end{cases}
$$
\item The total magnification is calculated as $\mu=\Sigma |\mu_i|$.

\end{itemize}
In Fig.\,\ref{fig:mag-orange1-3D}, we show an example of the magnification curve in two dimensional coordinate.


\section{Number of solutions}
In this appendix, we discuss the possible number of images generated by the lens system of a star and plasma. The planet is not included for simplicity.

Considering a star surrounding with plasma, in the one-dimensional case, the equation (\ref{eq:reduced-lenseq}) becomes
\be
y = x - \frac{1}{x} + \frac{B_0}{x} \frac{1}{|x|} \, , \quad {B_0}>0 \, .
\label{eq:lenseq-starwithplasma}
\ee

With given plasma strength $B_0$ and the source position $y$, the number of solutions $x$ of eq.(\ref{eq:lenseq-starwithplasma}) can be different. Qualitatively, this can be understood by plotting the right side of the eq.(\ref{eq:lenseq-starwithplasma}) as a function of $x$. Then each intersection of the resulting curves with the horizontal line $y=const$ will give us the solution to the eq.(\ref{eq:lenseq-starwithplasma}).

\begin{figure*}
\centerline{
\includegraphics[width=5.4cm]{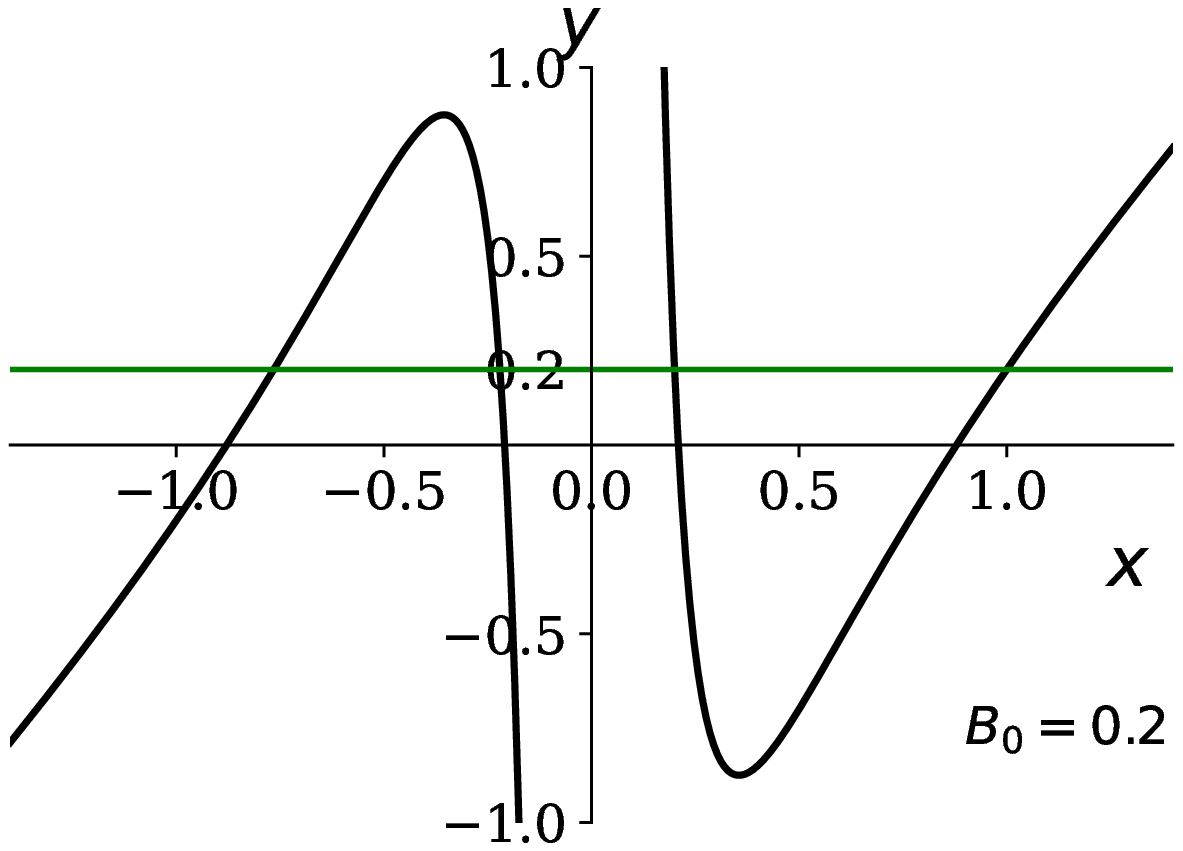}
\includegraphics[width=5.4cm]{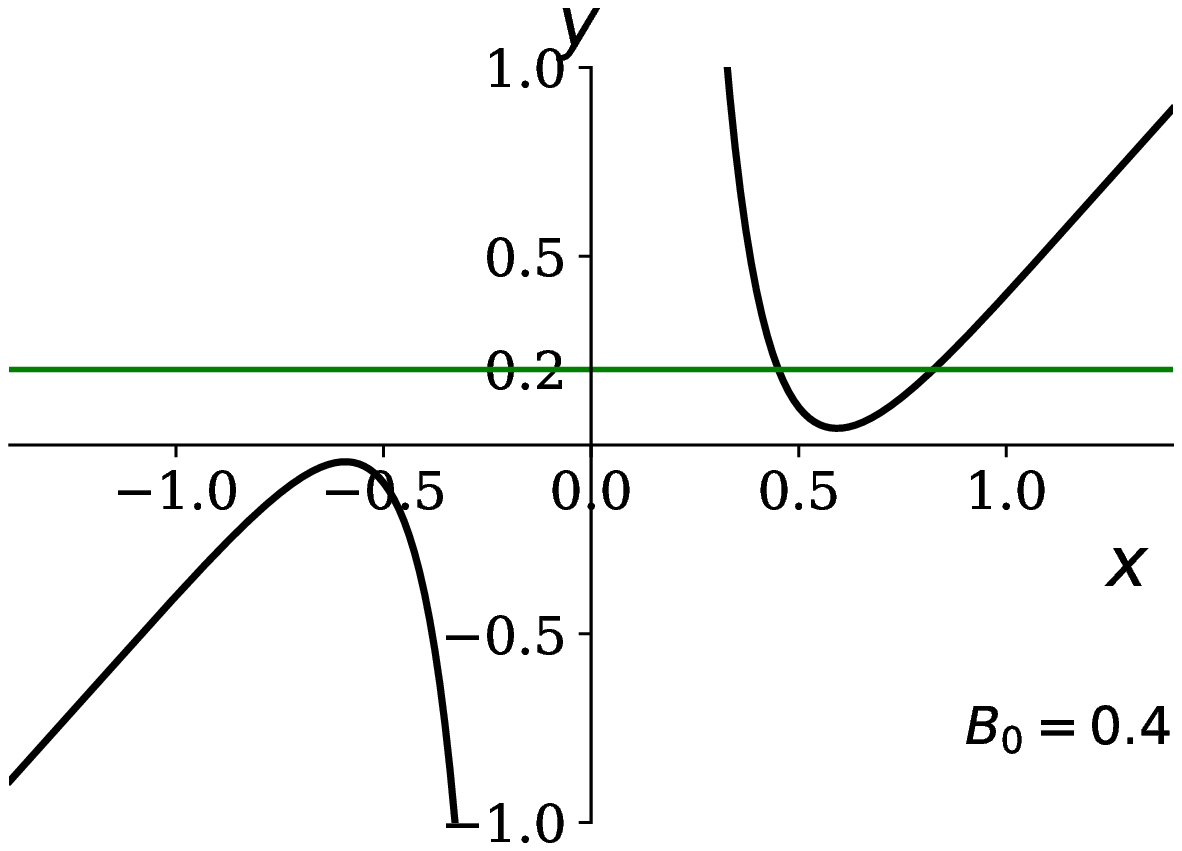}
\includegraphics[width=5.4cm]{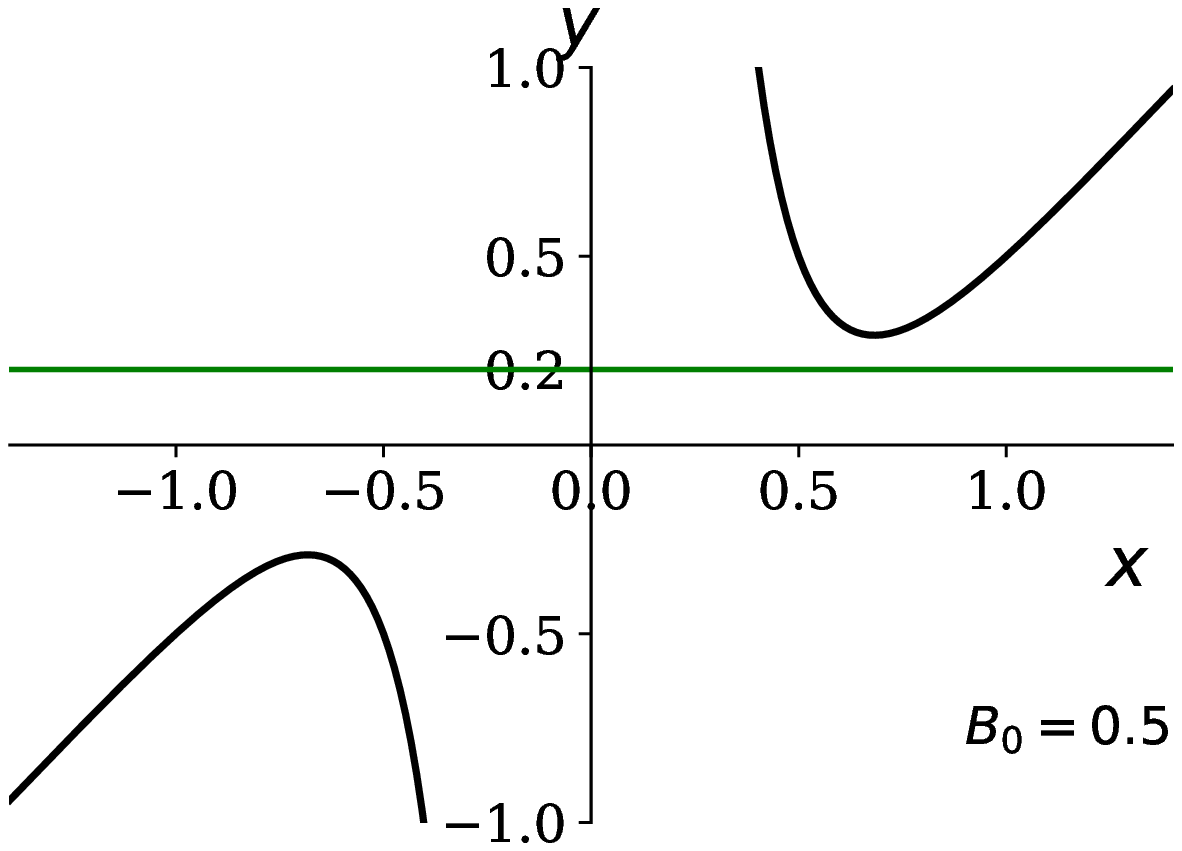}
}
\caption{The relationship between $x$ and $y$ with different $B_0$. If $y=0.2$, $B_0=0.2$ (left), thus $\Delta_1<0$, $\Delta_2<0$, there are four real solutions. If $y=0.2$, $B_0=0.4$ (median), thus $\Delta_1<0$, $\Delta_2>0$, there are two real solutions. If $y=0.2$, $B_0=0.5$ (right), thus $\Delta_1<0$, $\Delta_2<0$, there is no real solution.}
\label{fig:num}
\end{figure*}

In Fig.\ref{fig:num}, we present right hand side of eq.(\ref{eq:lenseq-starwithplasma}) for different values of $B_0$ and the same value of $y=0.2$. One can clearly see that there are four solutions on the left panel, and two solutions on the centre panel. On the right panel there are no solutions, which means that a hole is forming in the microlensing curve.

More strictly, the number of solutions of Eq.(\ref{eq:lenseq-starwithplasma}) can be found from the analysis of cubic equations. Eq.(\ref{eq:lenseq-starwithplasma}) can be rewritten in the following way:
\begin{numcases}{}
    x^3-y x^2-x+B_0=0 & \textrm{if $x>0$} \, ,
    \label{eq:alter-lenseq-starwithplasma-1}\\
    x^3-y x^2-x-B_0=0 & \textrm{if $x<0$}
    \label{eq:alter-lenseq-starwithplasma-2} \, .
\end{numcases}

According to Shengjin formulas \citep{shengjin}, a cubic equation of form $a x^3+b x^2+c x+d=0$ ($a\neq 0$, and $a,b,c,d$\,$\in\,\mathbb{R}$) has the discriminant
\be
\Delta\equiv B^2-4A C \, ,
\ee
where $A\equiv b^2-3a c \, , B\equiv b c-9a d \, , C\equiv c^2-3b d$. The solution can be estimated:
\begin{itemize}
\item 
If $A=B=0$, equation will have only one real triple root;
\item
If $\Delta>0$, equation will have one real root and a pair of conjugate imaginary roots;
\item
If $\Delta=0$, equation will have three real roots, including a double root;
\item
If $\Delta<0$, equation will have three different real roots.
\end{itemize}

Let the $\Delta$ of equation (\ref{eq:alter-lenseq-starwithplasma-1}) be $\Delta_1$, the $\Delta$ of equation (\ref{eq:alter-lenseq-starwithplasma-2}) be $\Delta_2$. The number of solutions to equation (\ref{eq:lenseq-starwithplasma}) can be described in the following ways:\\
(1) When $\Delta_1$ and $\Delta_2$ are negative real numbers, the relationship between $B_0$ and $y$ is:
$$
\begin{cases}
-12+81B_0^2-54B_0 y-3y^2-12B_0 y^3< 0 \, ,\\
-12+81B_0^2+54B_0 y-3y^2+12B_0 y^3< 0 \, .
\end{cases}
$$
The equation (\ref{eq:alter-lenseq-starwithplasma-1}) has three different real roots, including two positive roots, the equation (\ref{eq:alter-lenseq-starwithplasma-2}) has three different real roots, including two negative roots. Therefore, equation (\ref{eq:lenseq-starwithplasma}) has four real roots.\\
(2)When $\Delta_1=\Delta_2=0$, both equations (\ref{eq:alter-lenseq-starwithplasma-1}) and (\ref{eq:alter-lenseq-starwithplasma-2}) have one positive root and one negative root. Equation (\ref{eq:lenseq-starwithplasma}) has two real roots.\\ 
(3)When $\Delta_1$ is positive real number, but $\Delta_2$ is negative real number (or $\Delta_1$ is negative real number, $\Delta_2$ is positive real number), the relationship between $B_0$ and $y$ is:
$$
\begin{cases}
-12+81B_0^2-54B_0 y-3y^2-12B_0 y^3>0 \, ,\\
-12+81B_0^2+54B_0 y-3y^2+12B_0 y^3<0 \, ,
\end{cases} \, \mbox{or}
$$
$$
\begin{cases}
-12+81B_0^2-54B_0 y-3y^2-12B_0 y^3<0 \, ,\\
-12+81B_0^2+54B_0 y-3y^2+12B_0 y^3>0 \, .
\end{cases}
$$
The equation (\ref{eq:alter-lenseq-starwithplasma-1}) has no positive real roots (or has three different real roots, including two positive roots), the equation (\ref{eq:alter-lenseq-starwithplasma-2}) has three different real roots, including two negative roots (has no negative real roots). Therefore, equation (\ref{eq:lenseq-starwithplasma}) has two real roots. \\
(4) When $\Delta_1$ and $\Delta_2$ are positive real numbers, the relationship between $B_0$ and $y$ is:
$$
\begin{cases}
-12+81B_0^2-54B_0 y-3y^2-12B_0 y^3>0 \, ,\\
-12+81B_0^2+54B_0 y-3y^2+12B_0 y^3>0 \, .
\end{cases}
$$
The equation (\ref{eq:alter-lenseq-starwithplasma-1}) has no positive real roots, and the equation (\ref{eq:alter-lenseq-starwithplasma-2}) has no negative real roots. Therefore, equation (\ref{eq:lenseq-starwithplasma}) has no real roots.\\

\section{A Gallery of caustics}
\begin{figure*}
\centerline{
\includegraphics[width=4.5cm]{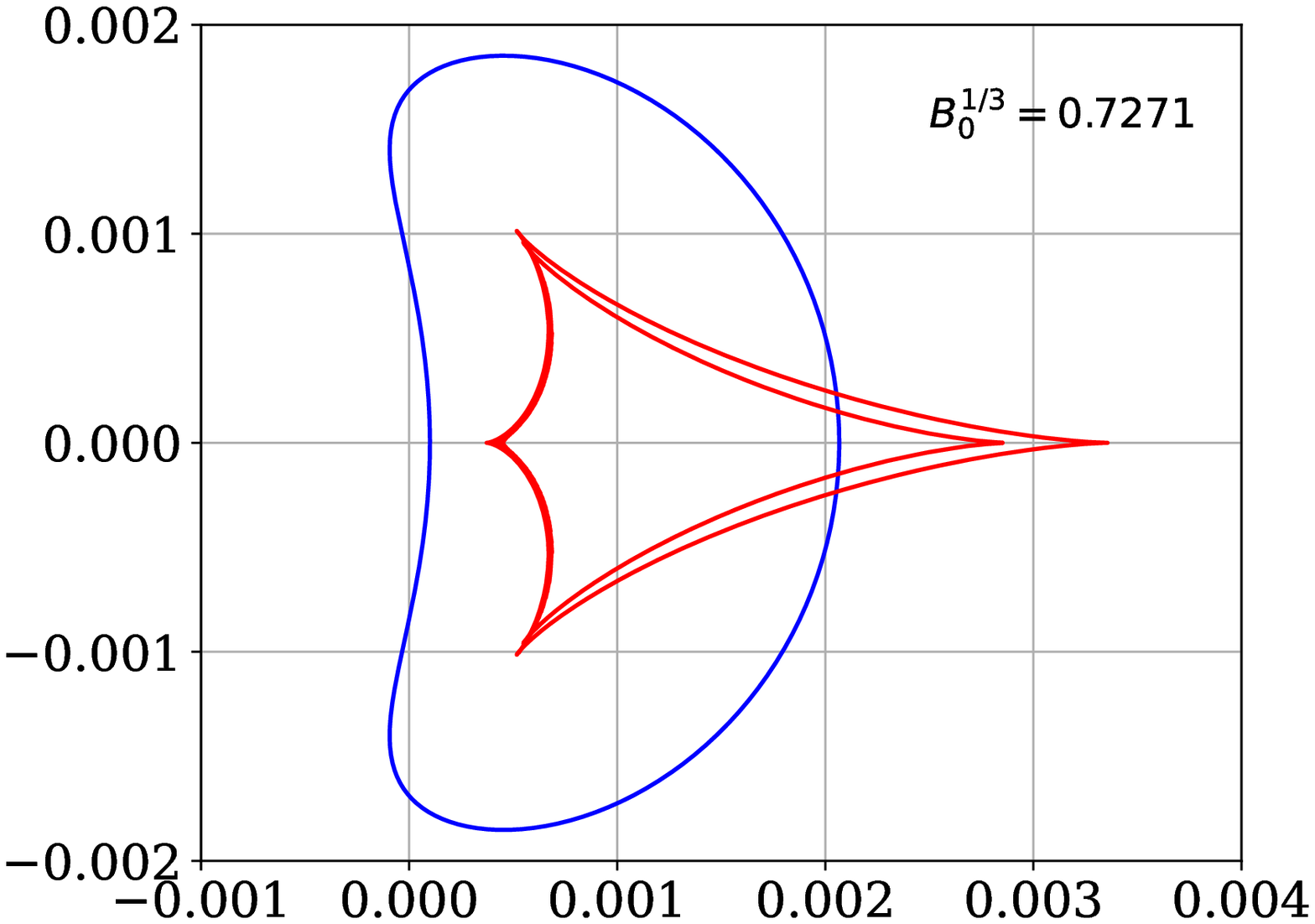}
\includegraphics[width=4.5cm]{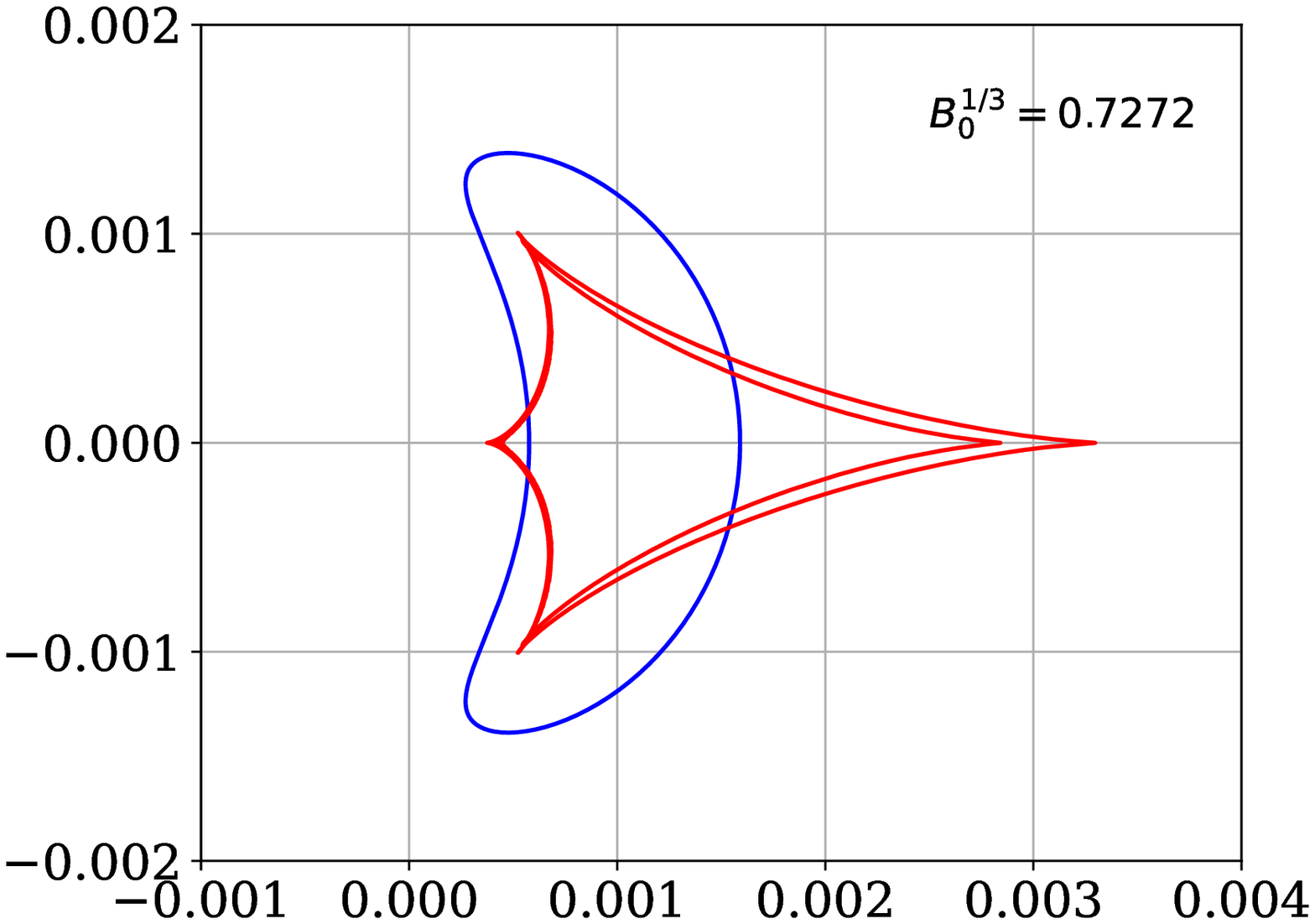}
\includegraphics[width=4.5cm]{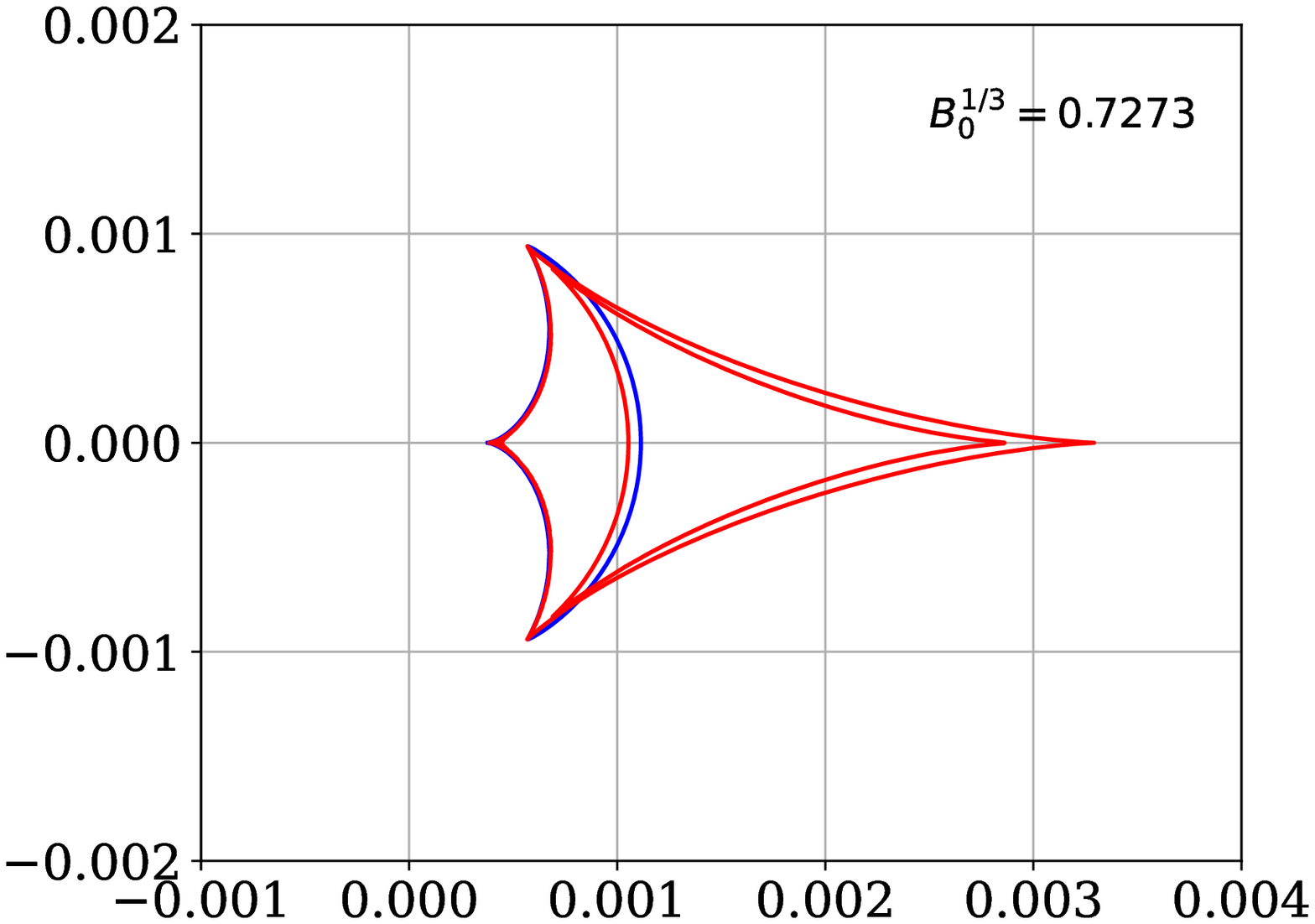}
\includegraphics[width=4.5cm]{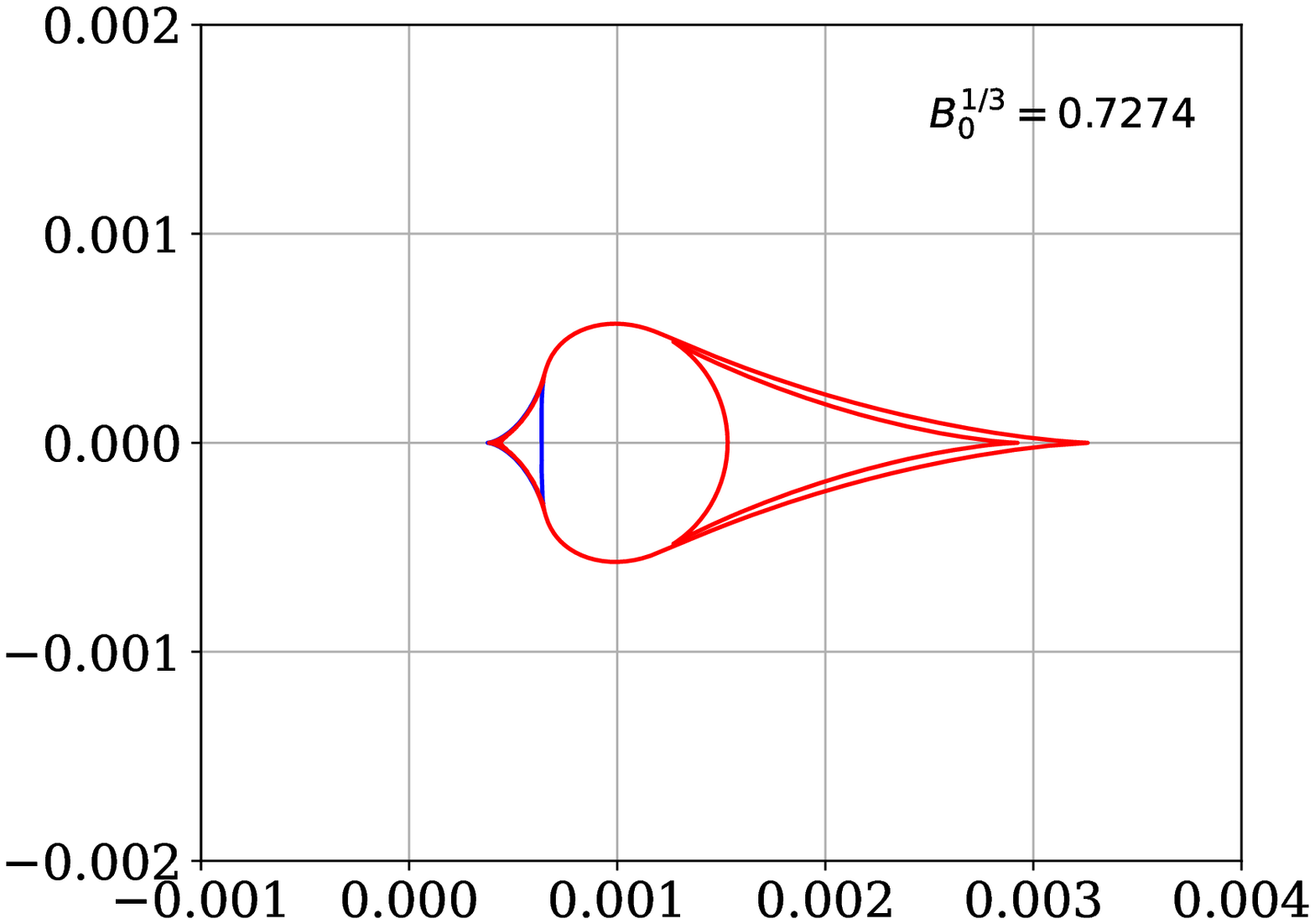}
}
\caption{The central caustics (red) and the extra caustics caused by plasma (blue) of binary lensing system. From left to right, we increase the density of plasma, which is indicated by $B_0$ in each panel. It can be seen that when $B_0^{1/3}=0.7273$, the blue line and red lines start to overlap. 
}
\label{fig:app-caustics7272}
\end{figure*}

\begin{figure*}
\centerline{
\includegraphics[width=4.5cm]{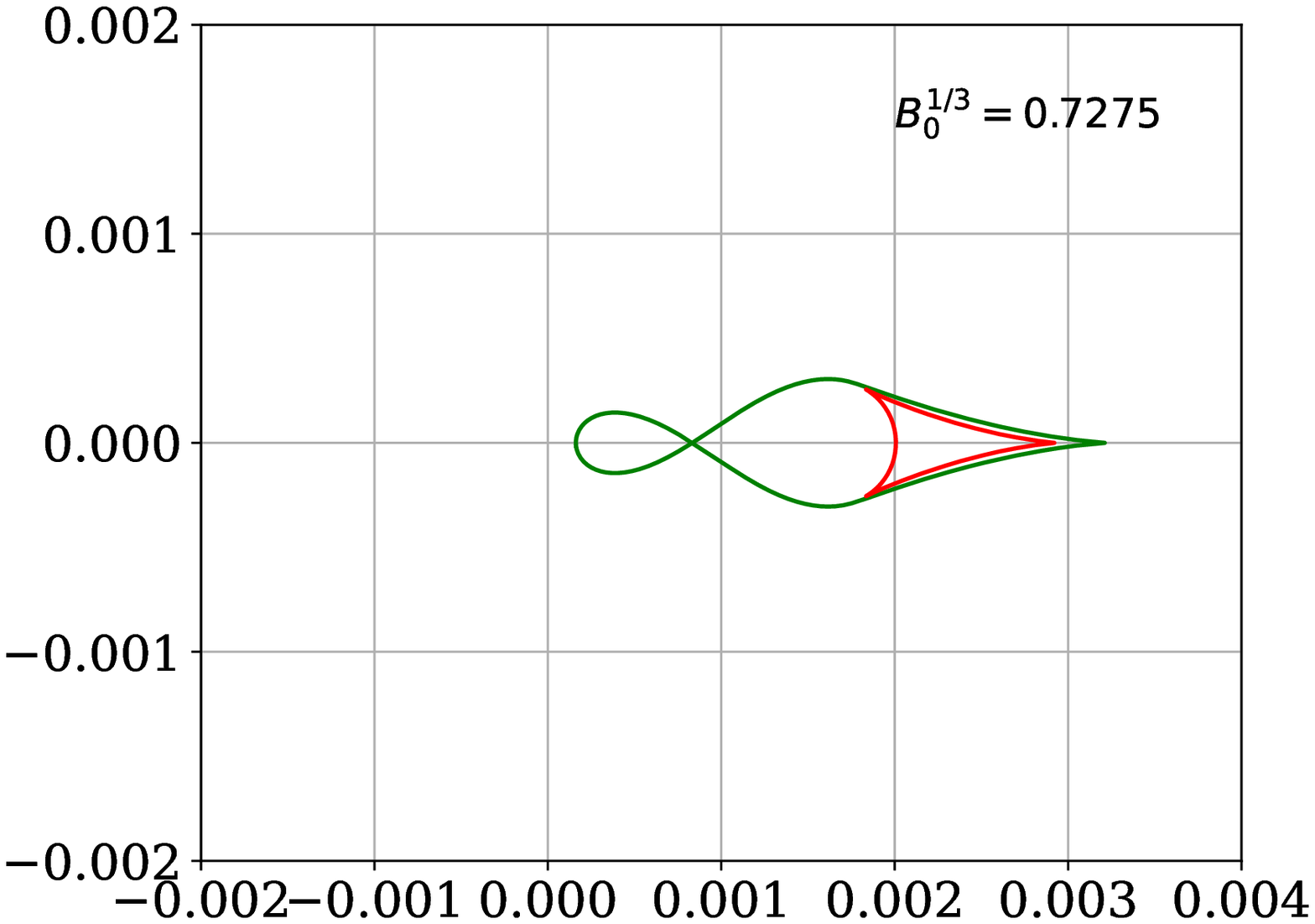}
\includegraphics[width=4.5cm]{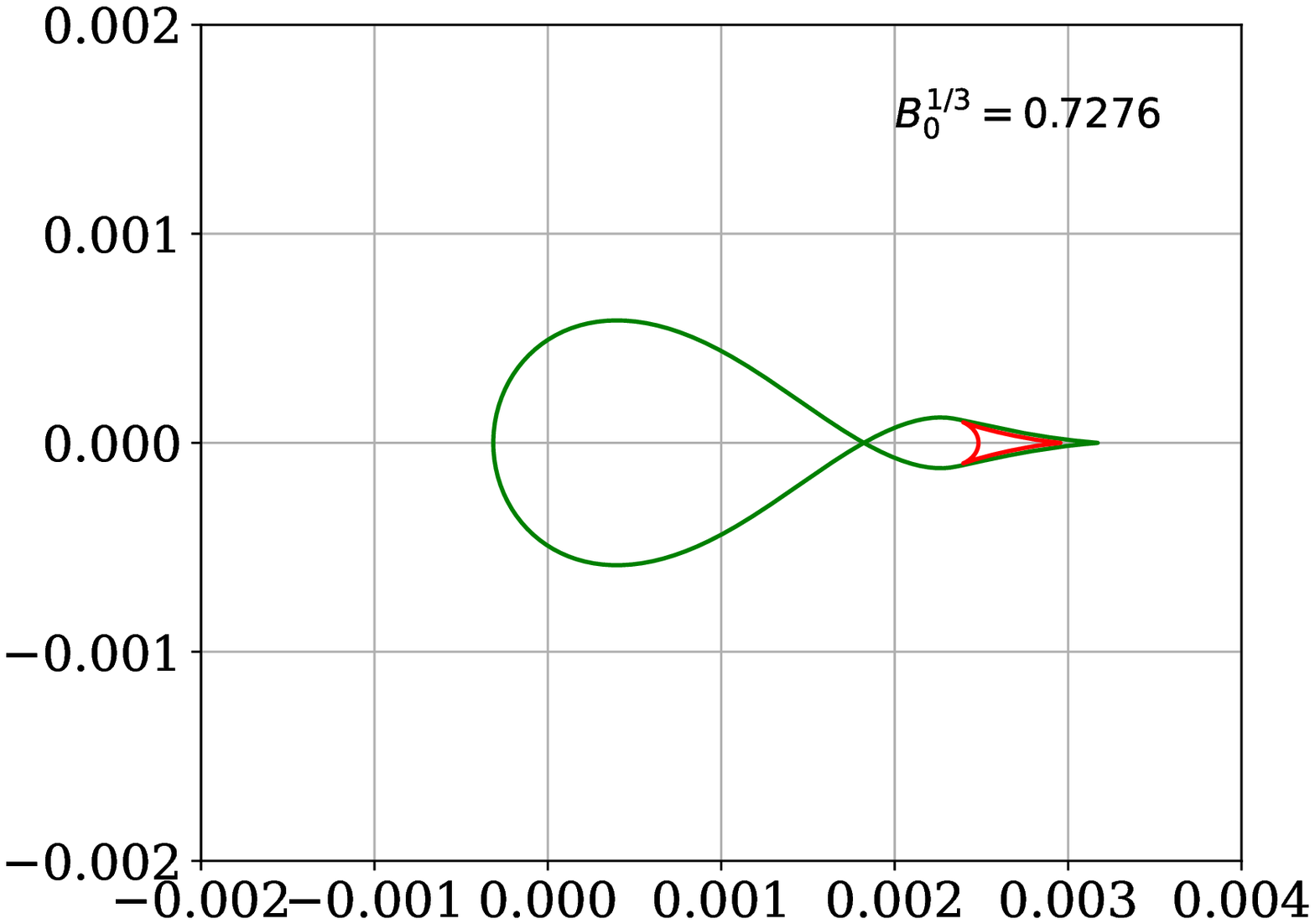}
\includegraphics[width=5cm]{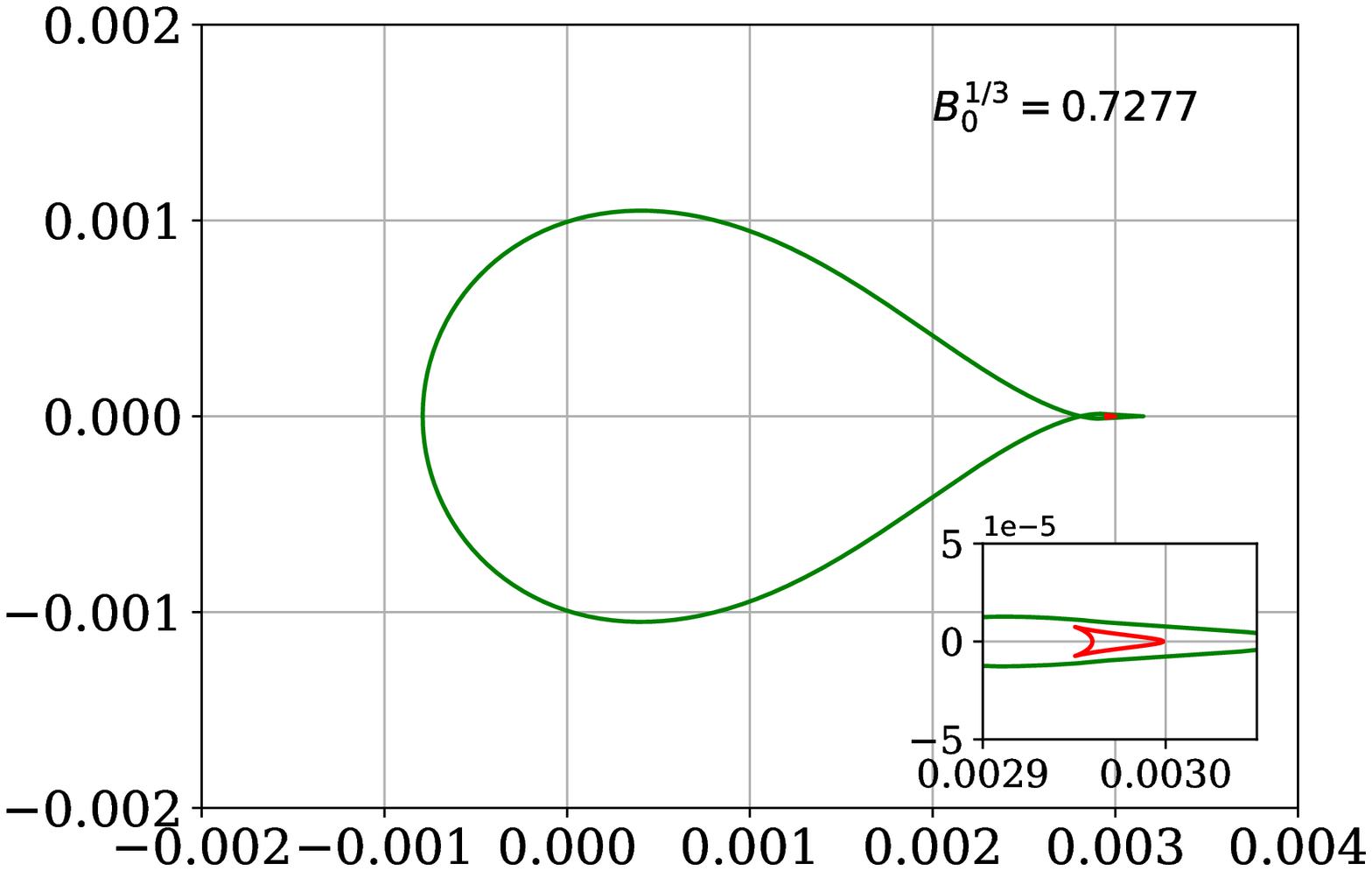}
\includegraphics[width=4.5cm]{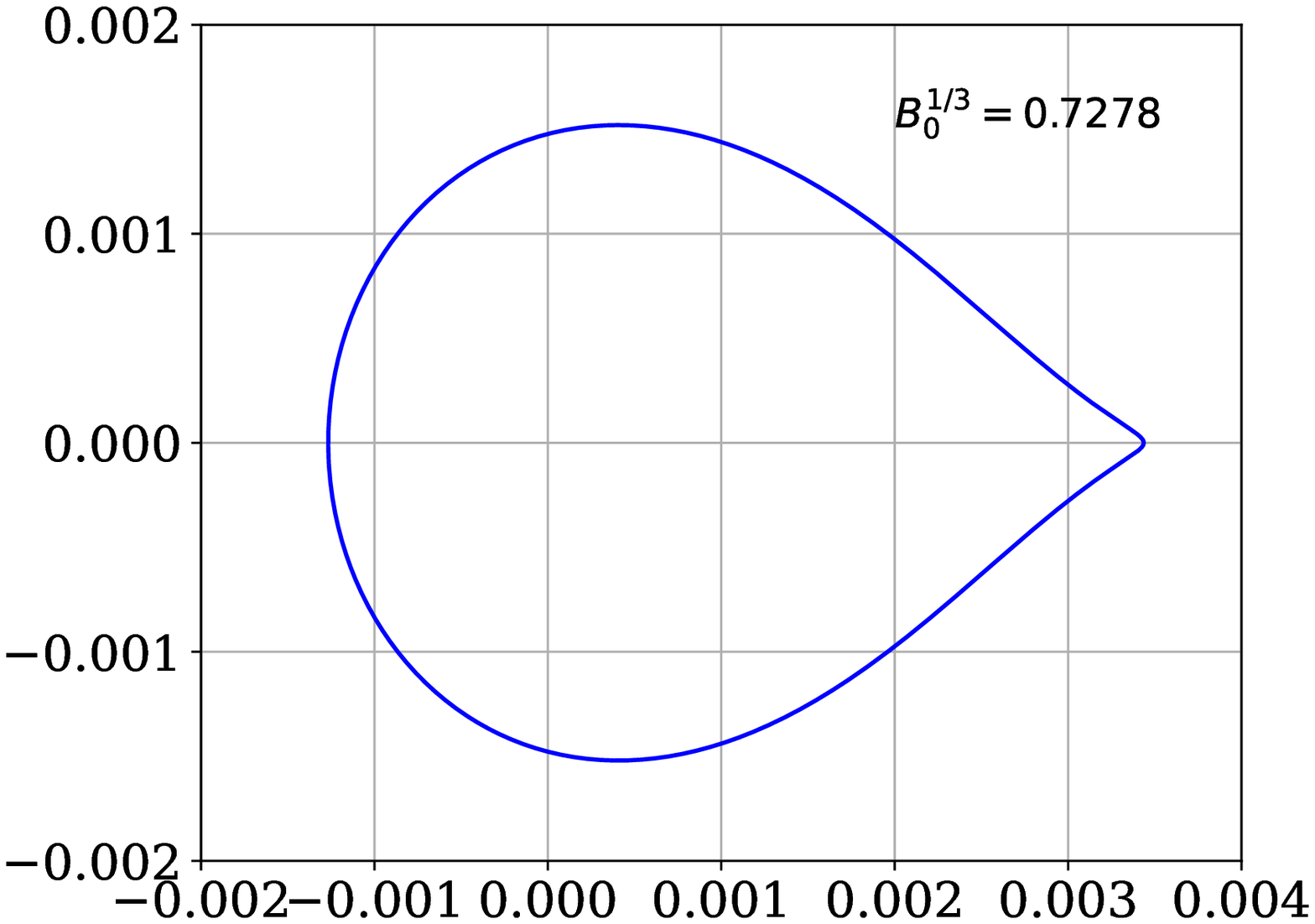}
}
\caption{When $B_0^{1/3}=0.7275$, the blue line and part of red lines completely integrate(green), and when $B_0^{1/3}=0.7278$, the influence of central caustics disappear.}
\label{fig:app-caustics7278}
\end{figure*}

In this appendix, we present the caustics for the binary system that have been discussed in Section 4. From Figs.\,\ref{fig:app-caustics7272} and \ref{fig:app-caustics7278}, one can see the transition of the caustics.

\section{'Hole' Formation}
\label{sec:hole-formation}
If planet is not considered, the critical value of $B_0$ for the 'hole' formation is \citep{2020MNRAS.491.5636T}:
\be
B_0=2 (h-1)^{\frac{h-1}{2}}/(h+1)^{\frac{h+1}{2}}.
\ee
So for $h=2$, the critical value is $B_0=2/3^{3/2}\approx 0.3849$.

\begin{figure*}
\includegraphics[width=5.5cm]{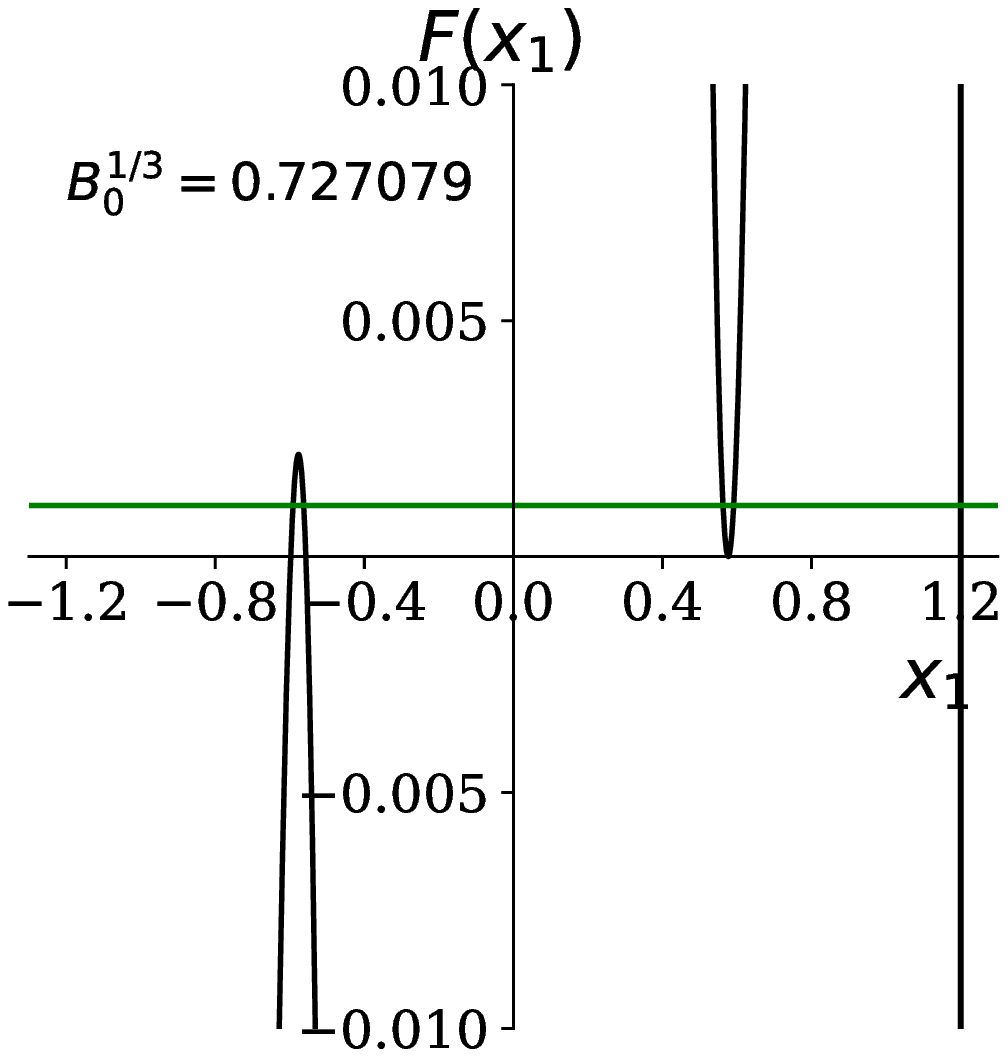}{1}
\includegraphics[width=5.5cm]{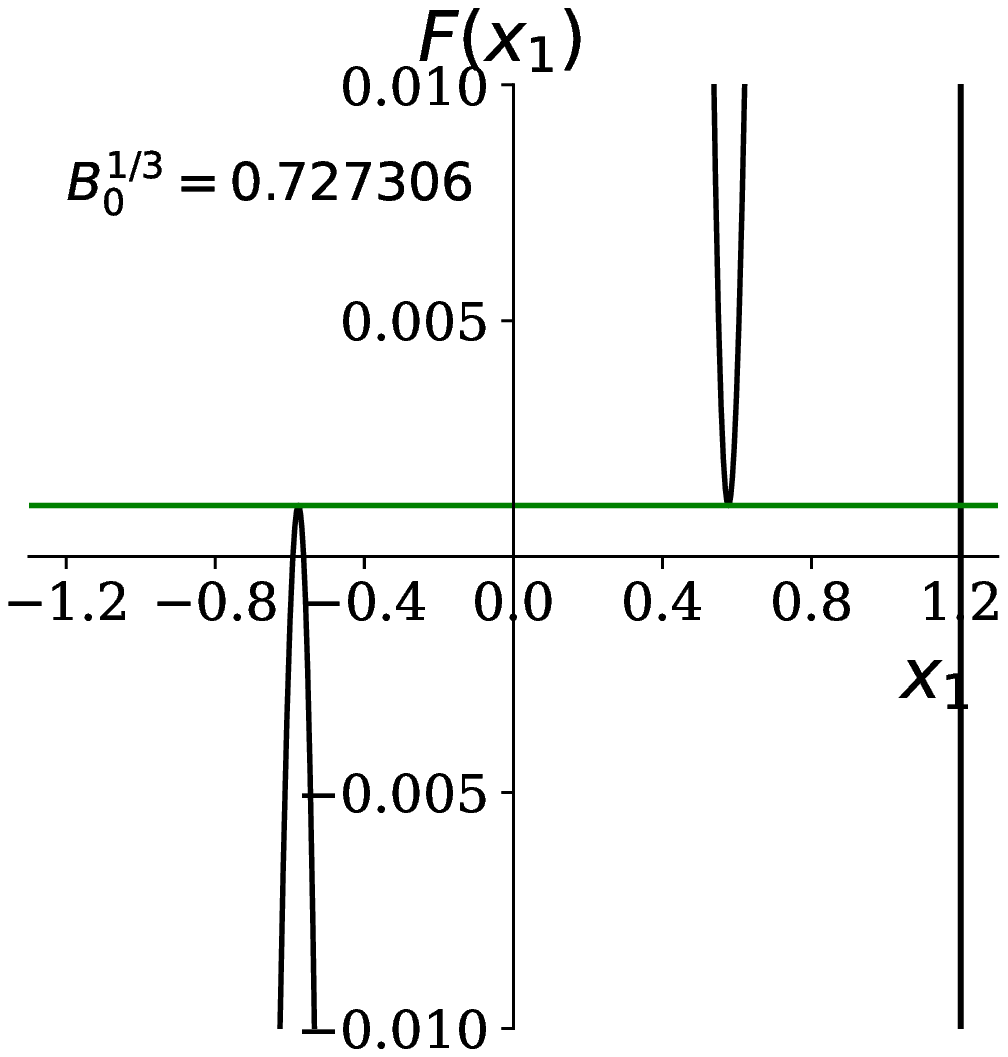}{2}
\includegraphics[width=5.5cm]{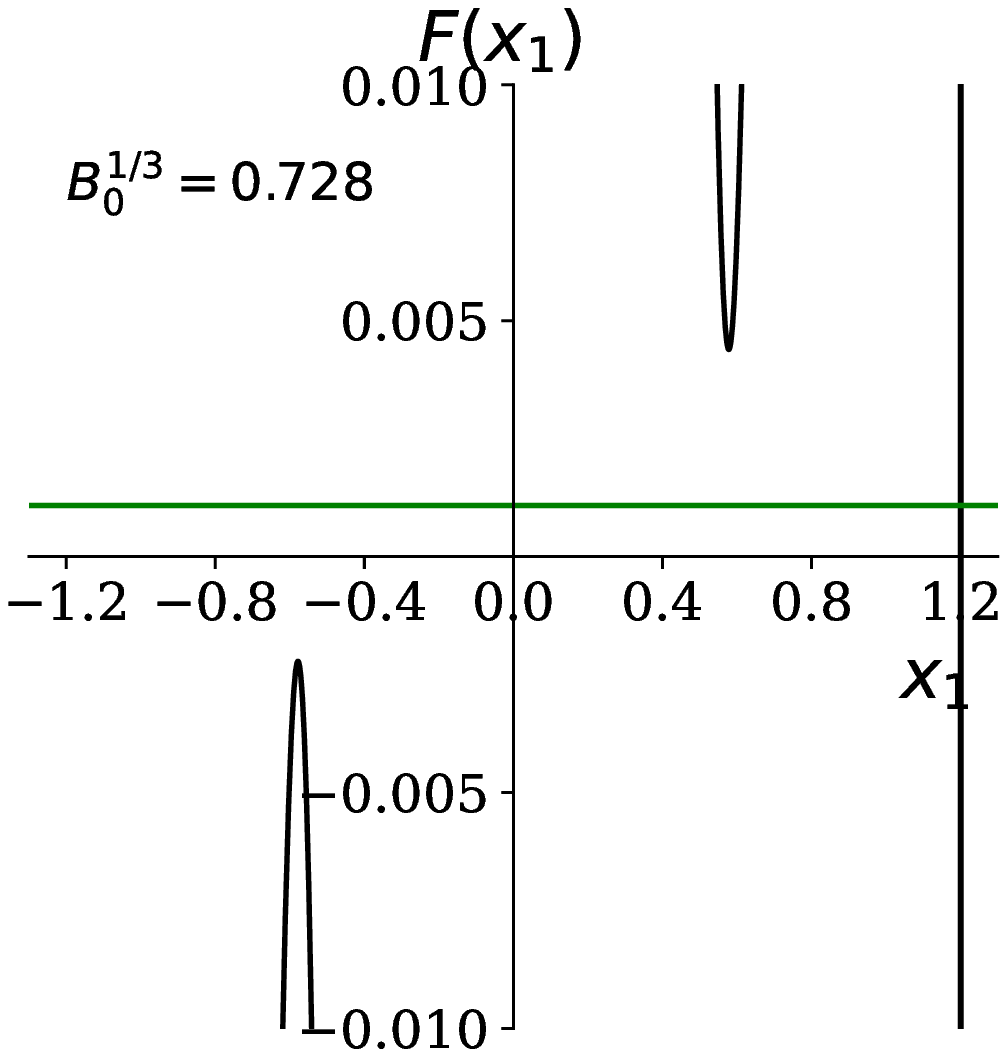}{3}
\caption{
The function $F(x_1)$ with different $B_0$. The green line is $F(x_1)$=0.001084, which indicates the position of the source where the central solution disappears.
}
\label{fig:app-hole}
\end{figure*}

If we consider the influence of planet, Eq.\,(\ref{eq:reduced-lenseq}) can be wrote as $F(x_1)$:
$$
F(x_1)=
\begin{cases}
x_1-\frac{1}{x_1}-\frac{0.001}{x_1-1.2}+\frac{B_0}{x_1^2}, \, x_1>0 \, ,\\ 
\\
x_1-\frac{1}{x_1}-\frac{0.001}{x_1-1.2}-\frac{B_0}{x_1^2}, \, x_1<0 \, .
\label{eq:F(x)_xithout_planet}
\end{cases}
$$
As in Fig.\,\ref{fig:app-hole}, when the local extremum on the right is 0, that on the left is not 0. Thus the critical value of $B_0$ does not correspond to the that when the right local minimum value equal to 0.

In order to calculate the critical value $B_0$, we can set the right local minimum at $x_{1_m}$, and the left local minimum at $x_{1_n}$, when $F(x_{1_m})=F(x_{1_n})$, the corresponding $B_0$ is the critical value, which equals to $0.384726$ approximately ($\theta_0/\theta_E=B_0^{1/3}\approx 0.727306$). In Fig.\,\ref{fig:app-hole}, we show the $F(x_1)$ with the critical value (panel 2), and the value smaller (panel 1) or greater (panel 3) than it. The horizontal line presents the position where the central image disappears. The interesting point is that with the inclusion of the planet, the central hole will form not at the exact ``centre'' of the lens. The intersection of the nearly vertical curve with the horizontal green line on the right side in each panel corresponds to the solution caused by the planet. 

\bsp	
\label{lastpage}
\end{document}